\documentclass[twocolumn,floats,floatfix,nofootinbib,11pt] {revtex4-1}
\usepackage{graphicx}
\usepackage{dcolumn}
\usepackage{bm}
\usepackage{graphics}
\usepackage{slashed}
\usepackage{amssymb}
\usepackage{natbib}
\usepackage{amsmath}
\usepackage{url}
\usepackage[usenames,dvipsnames,svgnames,table]{xcolor}
\usepackage{color}
\usepackage{xcolor}
\usepackage{tabularx}

\newcommand{\OO}{\mathcal{O}}

\begin{document}

\title{On the final fate of compact boson star mergers}

\author{
Miguel Bezares$^{1}$, %\footnote{Electronic address: vitor.cardoso@tecnico.ulisboa.pt},
Carlos Palenzuela$^{1}$,
Carles Bona$^{1}$
}

\affiliation{${^1}$Departament  de  F\'{\i}sica $\&$ IAC3,  Universitat  de  les  Illes  Balears  and  Institut  d'Estudis
Espacials  de  Catalunya,  Palma  de  Mallorca,  Baleares  E-07122,  Spain}

\begin{abstract}
Boson stars, self-gravitating objects made of a complex scalar field,
have been proposed as simple models for very different scenarios,
ranging from galaxy dark matter to black hole mimickers. 
Here we focus on a very compact type of boson stars to study binary mergers by varying different parameters, namely the phase shift, the direction of rotation and the angular momentum. Our aim is to investigate the properties of the object resulting from the merger in these different scenarios by means of numerical evolutions. These simulations, performed by using a modification of the covariant conformal Z4 (CCZ4) formalism of the Einstein Equations that does not require the algebraic enforcing of any constraint, indicate that the final state after a head-on collision of low mass boson stars is another boson star. However, almost complete annihilation of the stars occurs during the merger of a boson-antiboson pair. The merger of orbiting boson stars form a rotating bar that quickly relaxes to a non-rotating boson star.
\end{abstract}

\maketitle

% % % % % % % % % % % % % % % % % % % % % % % % % % % % % % % % % % % 
% % % % % % % % % % % % % % % % % % % % % % % % % % % % % % % % % % % 
\section{Introduction}
% % % % % % % % % % % % % % % % % % % % % % % % % % % % % % % % % % % 
% % % % % % % % % % % % % % % % % % % % % % % % % % % % % % % % % % % 

One of the most  prominent opportunities in the rising era of gravitational wave (GW) astronomy is to study  the strong-gravity regime through the signals produced during the coalescence of compact objects. Very recently, LIGO detectors observed the first GW signals consistent with binary black hole mergers~\cite{GW150914,GW151226},
although other sources could produce comparable waveforms~\cite{2016PhRvL.116q1101C,2017arXiv170101116C,car}. Consequently, these observations are already setting bounds both on the nature of these objects and on alternative theories of gravity~\cite{Yunes:2016jcc,TheLIGOScientific:2016src}.The next most likely candidate to be detected by GWs
is the coalescence of binary neutron stars. Neutron stars are very compact objects, usually possessing strong magnetic field, which existence has been confirmed with the observations of several binary pulsar systems on the electromagnetic (EM) band for several decades already (see for instance~\cite{doi:10.1146/annurev-nucl-102711-095018}). 

In addition to these known sources, gravitational waves can allow us to find unexpected astrophysical compact objects with low brightness, known generically as Exotic Compact Objects (ECOs). One of the most plausible ECO candidates are the boson stars (BSs), self-gravitating objects made of complex scalar field~\cite{liebpa,mace}. Even if their existence is still under debate due to the lack of any observational evidence, boson stars provide a simple and useful model to study compact bodies in very different scenarios, ranging from dark matter candidates to black hole mimickers. The maximum compactness $G M/c^2 R$ of the boson star depends on the self-interaction terms of its associated potential, and ranges from $\OO(10^{-3})$ for mini-boson stars to $\OO(10^{-1})$ for non-topological solitonic BS~\cite{frie,1992PhR...221..251L}.

Despite the simplicity of these smooth solutions, there are only few studies on binary boson star collisions within General Relativity.
Preliminary head-on collisions of mini-boson stars were first studied
in~\cite{1999PhDT........44B} within a 3D code. The dynamics of the merger, which showed an interesting interference pattern, was further analyzed  in~\cite{2004PhDT.......230L} with and axisymmetric code. Ultra-relativistic collisions were considered in~\cite{2010PhRvL.104k1101C}, and head-on and orbital mergers of non-identical boson stars  in~\cite{pale1,pale2}. Other related works include the study of the orbital case within the conformally flat approximation instead of full GR~\cite{2010PhDT.......390M}, and head-on collision of oscillatons~\cite{brito},  a solution analogous to BS but using just a real scalar field. Much more recently, collisions of solitonic boson stars has been numerically performed~\cite{car}, leading to dynamics qualitatively similar to the observed for mini-boson stars. 

This work aims to extend these recent numerical studies of solitonic boson stars by exploring a wider parameter space on the initial head-on configurations. We also consider binaries of identical solitonic boson stars with angular momentum to study whether the formation of a rotating boson star as a final state is possible. Our results indicate that, at least for the low mass boson stars considered here\footnote{The merger of high mass boson stars leads to the formation of a black hole~\cite{car}}, the head-on collision generically produce another boson star if the phase shift is not too close to $\pi$. However, the orbiting binary will not lead to a rotating boson star, but to a non-rotating one. This is probably due to the quantization of the angular momentum, and implies that the massive boson star formed by the merger must shed its excess of angular momentum by emitting GW and, in some cases, very rapid blobs of scalar field. 

Our simulations will be performed by using a novel modification of the CCZ4 formalism~\cite{alic} that treats all the constraints in the same manner. A common feature of the current conformal formulations, like the different flavors of BSSN\cite{1995PhRvD..52.5428S,BSSN} and CCZ4~\cite{bernu,alic}, is that a subset of the constraints of the system must be enforced after each time step of the simulation in order to obtain a stable evolution. Although this feature does not present a problem when using explicit time integrators, it might be not so straightforward for more sophisticated numerical methods or for automatically generated codes~\cite{Arbona20132321}. We introduce new terms in some of the equations to ensure that the full system is strongly hyperbolic (and well posed) and that the constraints are dynamically enforced during the evolution. Hence, our modified CCZ4 formalism does not require the algebraic enforcing of any constraint.

The paper is organized as follows. In section~\ref{ES} the Einstein-Klein-Gordon (EKG) evolution system is described in some detail. In particular, after a short summary on the CCZ4 formulation, we introduce novel modifications to avoid the algebraic enforcing of the constraints. In section~\ref{nts} the implementation of the evolution equations is briefly discussed, together with several numerical spacetimes --robust stability test, gauge waves and single solitonic boson star-- to test our evolution system. We study the dynamics of binary boson stars in Section~\ref{dsbs} by analyzing our numerical simulations of head-on and orbiting cases. Finally, we present our conclusion in section~\ref{con}. 
Throughout this paper, Roman letters from the beginning of the alphabet $a, b, c,...$ denote space-time indices ranging from 0 to 3, while letters near the middle $i, j, k,...$ range from 1 to 3, denoting spatial indices.
We also use geometric units in which $G = c = 1$, unless otherwise stated. 

% % % % % % % % % % % % % % % % % % % % % % % % % % % % % % % % % % % 
% % % % % % % % % % % % % % % % % % % % % % % % % % % % % % % % % % % 
\section{Evolution system }\label{ES}
% % % % % % % % % % % % % % % % % % % % % % % % % % % % % % % % % % % 
% % % % % % % % % % % % % % % % % % % % % % % % % % % % % % % % % % % 

The interaction between scalar-field matter and gravity, required to study the dynamics of boson stars, is given by the Einstein-Klein-Gordon equations. We adopt the CCZ4 formalism of the Einstein equations, which is briefly summarized next. We stress the modifications with respect to previous works and perform a characteristic analysis of the resulting system. The evolution equations for the complex scalar field are also described.

% % % % % % % % % % % % % % % % % % % % % % % % % % % % % % % % % % % 
\subsection{CCZ4 formalism}
% % % % % % % % % % % % % % % % % % % % % % % % % % % % % % % % % % % 

The Z4 formalism was first proposed as a covariant extension of Einstein equations to achieve an hyperbolic evolution system free of elliptic constraints~\cite{Z44,Z45}.
The equations of motion, which might also be derived from a Palatini-type variation~\cite{Z41}, are
\begin{eqnarray}
   R_{ab} &+& \nabla_a Z_b + \nabla_a Z_b   = 
   8\pi \, \left( T_{ab} - \frac{1}{2}g_{ab} \,tr T \right) \nonumber \\
   &+& \kappa_{z} \, \left(  n_a Z_b + n_b Z_a - g_{ab} n^c Z_c \right),
\label{Z4cov}
\end{eqnarray}
where $R_{ab}$ is the Ricci tensor associated to the spacetime metric $g_{ab}$, $T_{ab}$ is the stress-energy
tensor (with trace $tr T \equiv g^{ab} T_{ab}$)
and $Z_{a}$ is a new four-vector which measures the deviation from Einstein's solutions. Although the original formulation, corresponding to the choice $\kappa_{z}=0$, was completely covariant, additional damping terms were included to enforce a dynamical decay of the constraint violations associated to $Z_a$. As it is shown in~\cite{gundlach}, all the physical constraint modes are exponentially damped if $\kappa_{z} >0$. However, since the damping terms are proportional to the unit normal of the time slicing $n_{a}$, the full covariance of the system is lost due to the presence of a privileged time vector.
 
A conformal and covariant version of the $Z4$ (CCZ4) can be obtained from the 3+1 decomposition of the evolution equations by using conformal variables~\cite{alic} (i.e., see also~\cite{bernu} for other conformal but non-covariant Z4 formulations). Since this formulation is the starting point to our modifications, we first briefly summarize the derivation of the equations.

The first step involves writing the line element by using the 3+1 decomposition, namely
\begin{equation}
  ds^2 = - \alpha^2 \, dt^2 + \gamma_{ij} \bigl( dx^i + \beta^i dt \bigr) \bigl( dx^j + \beta^j dt \bigr), 
\label{3+1decom}  
\end{equation}
where $\alpha$ is the lapse function, $\beta^{i}$ is the shift vector and $\gamma_{ij}$ is the induced metric in each spatial foliation. In this foliation geometry we can define
the normal to the hypersurfaces $n_{a}=(-\alpha,0)$
%, so that $n^{a}=(1/\alpha,- \beta^{i}/\alpha)$. 
%We can also define the extrinsic curvature $K_{ij}
and the extrinsic curvature $K_{ij} \equiv  -\frac{1}{2}\mathcal{L}_{n}\gamma_{ij}$,  where $\mathcal{L}_{n}$ is the Lie derivative along  $n^{a}$. Therefore, the Z4 formalism given by eq.~(\ref{Z4cov}), together with the metric decomposition eq.~(\ref{3+1decom}) and these definitions, lead to evolution equations for the evolved fields
$\{ \gamma_{ij}, K_{ij}, Z_i, \Theta  \}$, where
we have defined $\Theta \equiv - n_{a} Z^{a}$.

In the second step a conformal decomposition is applied to the evolved fields. A conformal metric ${\tilde \gamma}_{ij}$ with unit determinant and a conformal trace-less extrinsic curvature ${\tilde A}_{ij}$ can be defined as 
\begin{eqnarray}
  {\tilde \gamma}_{ij} &=& \chi \,\gamma_{ij} , \\
  {\tilde A}_{ij} &=& \chi \, \bigl( K_{ij} - {1\over3} \, \gamma_{ij} \, tr K \bigr),
\end{eqnarray} 
where $tr K \equiv \gamma^{ij}K_{ij}$.
These definitions lead to the following new constraints 
\begin{eqnarray}
  {\tilde \gamma} = 1 ~~~~~~ , ~~~~~~
  {tr \tilde A} \equiv {\tilde \gamma}^{ij} {\tilde A}_{ij}  = 0,
  \label{enf}
\end{eqnarray} 
which will be denoted as {\em conformal constraints} from now on
to distinguish them from the {\em physical constraints} associated to $Z_a$.
Notice that now the evolved fields are
$\{ \chi, {\tilde \gamma}_{ij}, trK, {\tilde A}_{ij}, Z_i, \Theta  \}$.

Instead of using $trK$ and $Z_i$, it is more convenient to use the following quantities
\begin{eqnarray}
 {tr \hat K} &\equiv& tr K - 2\, \Theta, \\
{\hat \Gamma}^i &\equiv& {\tilde \Gamma}^i + {2 \over \chi}  Z^{i} , 
\end{eqnarray}
so that the evolution equations are closer to those in the BSSN formulation \cite{1995PhRvD..52.5428S,BSSN}, where the quantity ${\tilde \Gamma}^i = {\tilde \gamma}^{jk} \, {\tilde \Gamma}^i{}_{jk} = - \partial_j {\tilde \gamma}^{ij}$ is directly evolved. Therefore, the final list of evolved fields become
$\{ \chi, {\tilde \gamma}_{ij}, {tr \hat K}, {\tilde A}_{ij}, {\hat \Gamma}^i, \Theta  \}$, following the evolution equations 
\begin{widetext}
\begin{eqnarray}
\partial_t {\tilde \gamma}_{ij} 
    & =& \beta^k \partial_k {\tilde \gamma}_{ij} + {\tilde \gamma}_{ik} \, \partial_j \beta^k 
    + {\tilde \gamma}_{kj} \partial_i \beta^k - {2\over3} \, {\tilde \gamma}_{ij} \partial_k \beta^k
    - 2 \alpha \Bigl( {\tilde A}_{ij}  - \frac{\lambda_{0}}{3} {\tilde \gamma}_{ij}\, tr {\tilde A} \Bigr) -  \frac{\kappa_{c}}{3}\,\alpha\tilde{\gamma}_{ij}\ln\tilde{\gamma}, \label{syseq1}
\\
\partial_t {\tilde A}_{ij} 
    & =& \beta^k \partial_k{\tilde A}_{ij} + {\tilde A}_{ik} \partial_j \beta^k 
     + {\tilde A}_{kj} \partial_i \beta^k - {2\over3} \, {\tilde A}_{ij} \partial_k \beta^k - \,\frac{\kappa_{c}}{3}\,\alpha\,\tilde{\gamma}_{ij}
     \,tr \tilde{A}
\\
& +& \chi \, \Bigl[ \, \alpha \, \bigl( {^{(3)\!}R}_{ij} + \nabla_i Z_j + \nabla_j Z_i 
    - 8 \pi \, S_{ij} \bigr)  - \nabla_i \nabla_j \alpha \, \Bigr]^{\rm TF} 
 + \alpha \, \Bigl( tr {\hat K} \, {\tilde A}_{ij} - 2 {\tilde A}_{ik} {\tilde A}^k{}_j \Bigr), 
\nonumber \\
\partial_t \chi & =& \beta^k \partial_k \chi 
+ {2\over 3} \, \chi \, \bigl[ \alpha (tr {\hat K} + 2\, \Theta) - \partial_k \beta^k  \bigr], 
\\
\partial_t tr {\hat K} 
    & =&  \beta^k \partial_k tr {\hat K} 
       - \nabla_i \nabla^i \alpha
        + \alpha \, \left[ {1 \over 3} \bigl( tr {\hat K} + 2 \Theta \bigr)^2 
       + {\tilde A}_{ij} {\tilde A}^{ij} + 4\pi  \bigl(\tau + tr S\bigr)
       + \kappa_z  \Theta \right]  
\nonumber \\
       &+& 2\, Z^i \nabla_i \alpha, 
\\
\partial_t \Theta 
    & =&  \beta^k \partial_k \Theta + {\alpha \over 2} \left[ {^{(3)\!}R} + 2 \nabla_i Z^i
   + {2\over3} \, tr^2{\hat K} + {2\over3} \, \Theta \Bigl( tr {\hat K} - 2 \Theta \Bigr)
   - {\tilde A}_{ij} {\tilde A}^{ij}  \right] - Z^i \nabla_i \alpha 
\nonumber \\
   &-& \alpha \, \Bigl[ 8\pi  \, \tau  + 2\kappa_z  \, \Theta \Bigr], 
\\   
\partial_t {\hat \Gamma}^i 
    & =& \beta^j \partial_j {\hat \Gamma}^i - {\hat \Gamma}^j \partial_j \beta^i 
    + {2\over3} {\hat \Gamma}^i \partial_j \beta^j + {\tilde \gamma}^{jk} \partial_j \partial_k \beta^i
   + {1\over3} \, {\tilde \gamma}^{ij} \partial_j \partial_k \beta^k \nonumber
\\
& -& 2 {\tilde A}^{ij} \partial_j \alpha + 2\alpha \, \Bigl[ {\tilde \Gamma}^i{}_{jk} {\tilde A}^{jk}
  - {3 \over 2 \chi} \, {\tilde A}^{ij} \partial_j \chi 
  - {2\over3} \, {\tilde \gamma}^{ij} \partial_j tr{\hat K} - 8\pi \, {\tilde \gamma}^{ij} \, S_i \Bigr] \nonumber
\\ 
& +& 2 \alpha \, \left[- {\tilde \gamma}^{ij} \left( {1 \over 3}\partial_j \Theta 
  + {\Theta \over \alpha} \, \partial_j \alpha \right) 
  - {1 \over \chi} Z^i \left( \kappa_z + {2\over 3} \, (tr{\hat K} + 2 \Theta) \right) \right],   
   \label{syseq2}
\end{eqnarray}
\end{widetext}
where the expression $[\ldots]^{\rm TF}$ indicates the trace-free part with respect to the metric $\tilde{\gamma}_{ij}$. 
The non-trivial terms inside this expression can be written as 
\begin{eqnarray}
{^{(3)\!}R}_{ij} & + & 2  \nabla_{(i} Z_{j)} 
   = {^{(3)\!}{\hat R}}_{ij} + {\hat R}^\chi_{ij} ,
\nonumber \\
{\hat R}^{\chi}_{ij} & = & {1 \over 2 \chi} \, \partial_i \partial_j \chi 
    - {1 \over 2 \chi} \, {{\tilde \Gamma}^k}_{ij} \partial_k \chi 
\nonumber \\    
   & -& {1 \over 4 \chi^2} \, \partial_i \chi \partial_j \chi 
    + {2 \over \chi^2} Z^k {\tilde \gamma}_{k(i} \partial_{j)} \chi 
\nonumber\\
 &+& {1 \over 2 \chi }{\tilde \gamma}_{ij} \, \Bigl[ {\tilde \gamma}^{km} \Bigl( {\partial}_k {\partial}_m \chi 
 -  {3\over 2 \chi} \, \partial_k \chi \partial_m \chi \Bigr)
\nonumber \\ 
 & &~~~~~~ - {\hat \Gamma}^k \partial_k \chi \Bigr] ,
\nonumber     \\
{^{(3)\!}{\hat R}}_{ij} &=& - {1\over2} \, {\tilde \gamma}^{mn} \partial_m \partial_n {\tilde \gamma}_{ij}     
        + {\tilde \gamma}_{k(i} \partial_{j)} {\hat \Gamma}^k 
\nonumber
\\        
        &+&  {\hat \Gamma }^k {\tilde \Gamma}_{(ij)k} + {\tilde \gamma}^{mn} 
             \Bigl(  {{\tilde \Gamma}^k}_{mi} {\tilde \Gamma}_{jkn} 
        \Bigr.
\nonumber   \\
         &+&  {{\tilde \Gamma}^k}_{mj} {\tilde \Gamma}_{ikn} +{\tilde \Gamma}^k{}_{mi} {\tilde \Gamma}_{knj}  \Bigr), 
\nonumber \\
 \nabla_i \nabla_j \alpha & = & \partial_i \partial_j \alpha - {{\tilde \Gamma}^k}_{ij} \partial_k \alpha + {1 \over 2 \chi} \Bigl( \partial_i \alpha\, \partial_j \chi \Bigr.
\nonumber \\ 
  &+&  \Bigl. \partial_j \alpha  \, \partial_i \chi
           - {\tilde \gamma}_{ij}\, {\tilde \gamma}^{km}\, \partial_k \alpha\, \partial_m \chi  \Bigr),  
\nonumber
\end{eqnarray}

The matter terms are computed
by contracting the stress-energy tensor, namely 
\begin{eqnarray}
   \tau &=&  n_a \, n_b \, T^{ab} = \alpha^2 \, T^{00}, \label{matter1}
\\
   S_i &=& -n^a T_{ai} \equiv {{\tilde S}_i \over \chi}
~,~
   S_{ij} =  T_{ij} \equiv {{\tilde S}_{ij} \over \chi^2},
\\
 {\tilde S}_i &=& \alpha\, {\tilde \gamma}_{ik} \Bigl( T^{0k} + \beta^k\, T^{00}\Bigr),
\\ 
{\tilde S}_{ij} &=& {\tilde \gamma}_{ik} \, {\tilde \gamma}_{jm} \, \beta^k \beta^m \, T^{00} 
       + \Bigl( {\tilde \gamma}_{ik}\, \beta_j + {\tilde \gamma}_{jk}\, \beta_i \Bigr) T^{0k}\nonumber \\
       &+& {\tilde \gamma}_{ik}\, {\tilde \gamma}_{jm}\, T^{km} , \label{matter2}
\end{eqnarray}

%Actually, the original work \cite{alic} there are three damping terms but for our purpose we only consider $ \kappa_{1} $ damping term and $\kappa_{2}=\kappa_{3}=0$.

The evolution equations (\ref{syseq1}-\ref{syseq2}) are 
equivalent to those obtained in~\cite{alic}, by defining the conformal
factor $\chi = \gamma^{-1/3}$ instead of $\phi=\gamma^{-1/6}$,
except by two significant differences. First, there is a new term proportional to $tr\tilde{A}$. This term, which was already suggested in~\cite{alic}, is crucial to obtain a
well-posed evolution system if the algebraic conformal constraints $\ln {\tilde \gamma}= tr {\tilde A} = 0$ are not enforced during the evolution. Second, damping terms proportional to $\kappa_c$ have been included in order to dynamically control the conformal constraints, exactly in the same way as it is done with the physical ones.

In order to close the system of equations, coordinate (or gauge) conditions for the evolution of the lapse and shift must be supplied. We use the Bona-Mass\'o family of slicing conditions~\cite{BM} and the Gamma-driver shift condition~\cite{alcub}, namely 
\begin{eqnarray}
\partial_t \alpha & = &  \beta^i \partial_i \alpha - \alpha^{2} \,f\, tr\hat{K}, \\ 
\partial_t \beta^i & = &  \beta^j \partial_j \beta^i +  \,g \, \, B^{i}, \\
\partial_t B^i & = & \beta^j \partial_j B^i - \eta B^i %\nonumber \\
 + \partial_t {\hat \Gamma}^i - \beta^j \partial_j {\hat \Gamma}^{i},
\end{eqnarray}
being $f$ and $g$ arbitrary functions depending on the lapse and the metric, and $\eta$ a constant parameter.

\subsection{Characteristic structure}\label{hyperCCZ4}

Now that the system is complete, it is possible to calculate its
characteristic structure. Here we use the concept of pseudo-hyperbolicity~\cite{Z43,CCC}, that relies on a plane-wave analysis applied to the linearized equations around a background metric \footnote{Notice also the work in Ref.~\cite{nagy} that extend these ideas using pseudo-differential operators.}.
Hence, we consider the line element 
\begin{equation}
ds^2 = -\alpha_{0}^{2}dt^{2}  + \tilde{\gamma}^{0}_{ij}(dx^{i} + \beta_{0}^{i}dt)(dx^{j} + \beta_{0}^{j}dt),
\end{equation}
and study the dynamics of perturbations over this background spacetime which propagates along a given normalized direction $s_{i}$ (i.e., such that $\gamma_{ij}^0 s^{i} s^{j}= 1$). The perturbation for the metric fields $\{\alpha, \beta^{k}, \tilde{\gamma}_{ij}, \chi\}$ has a plane-wave form,
\begin{equation}
   g_{ab} - g_{ab}^0  = e^{i \omega_k x^k }\,\bar{g}_{ab}(\omega,t) ,
\end{equation}
where $\omega^k$ is the wavenumber and $\omega \equiv \omega_{k}\,s^{k}$. An additional factor $i \omega$ appears in the perturbations of the fields
$\{\tilde{A}_{ij}, \hat{K}, \Theta, \hat{\Gamma}^{i},  B^{i} \}$, which are first derivatives of the metric, namely
\begin{equation}
  K_{ab}  = i\,\omega\, e^{i\omega_k x^k}\,\bar{K}_{ab}(\omega,t),
\end{equation}
Replacing the above mentioned definitions in (\ref{syseq1}\,-\,\ref{syseq2}) one can obtain the following system:
\begin{equation}
  \partial_{t} \bar{u} = -i\,\omega(\mathcal{A} - \bar{\beta}^{s}_{0}\mathrm{I})\bar{u},
  \label{eqgen}
\end{equation}

where $\bar{u}$ is a vector containing the perturbation of the fields, $\mathcal{A}$ is the characteristic matrix and
$\mathrm{I}$ the identity one. The index $s$ means a contraction with the propagation direction $s_{i}$ (i.e.,   $\bar{\beta}^{s}_{0} = s_{i}\bar{\beta}_{0}^{i}$). The projection orthogonal to $s_{i}$ will be denoted by the index $\perp$. 

The system~\eqref{eqgen} is pseudo-hyperbolic if and only if  the characteristic matrix $\mathcal{A}$ has real eigenvalues and a complete set of eigenvectors. Therefore, with this definition, hyperbolicity of the system translates into a set of algebraic conditions~\cite{reula}. The analysis of the  characteristic structure can be simplified by splitting the perturbations in different sectors which do not interact (i.e., or at least, not strongly) with the others.

It is instructive to analyze first the effect of the term proportional to $\lambda_0$.
There is a sector, involving only the perturbations of $\tilde{\gamma}$ and $tr \tilde{A}$, given by:
\begin{equation}
\partial_{t}
\begin{pmatrix}
\tilde{\gamma} \\
tr \tilde{A}
\end{pmatrix} \approx \alpha
\begin{pmatrix}
0 & 2 (\lambda_{0}-1) \\
0 & 0 \\
\end{pmatrix}
\begin{pmatrix}
\tilde{\gamma} \\
tr \tilde{A}
\end{pmatrix} + \ldots ,
\end{equation}
where $\approx$ means that only the principal part is considered. Obviously, for the original choice $\lambda_{0}=0$, there is not a complete set of eigenvectors. Then, the system is only weakly pseudo-hyperbolic system and, consequently, the problem is ill-posed. The same problem appears for any other value except for $\lambda_{0}=1$. Only for this choice both $\{\tilde{\gamma},\tilde{A}\}$ are standing modes, implying that this sector has a complete set of eigenvectors. As it is shown next, the
other sectors are also complete, meaning that the full system is strongly pseudo-hyperbolic. 
Notice that this lack of strong hyperbolicity (together with the
unbound growth of the conformal constraints) prevents to evolve directly the unconstrained CCZ4, unless the conformal constraints are algebraically enforced during the evolution~\cite{hilditch}.

We can now study the characteristic structure of the other modes. The lapse sector, constituted by $\{ \bar{\alpha},tr \bar{K} \}$, has a complete set of eigenvectors with eigenvalues $-\bar{\beta}^{s}_{0}\pm \alpha_{0} \sqrt{f}$. The longitudinal shift and energy modes form another closed sector with a complete set of eigenvectors including  $\{\bar{\chi},\bar{\Theta},\bar{\Gamma}^{s},\bar{\beta}^{s},\bar{B}^{s}\}$ with characteristic speeds given by $\{-\bar{\beta}^{s}_{0},-\bar{\beta}^{s}_{0} \pm \sqrt{4\,g/3\chi_{0}}, -\bar{\beta}^{s}_{0} \pm \alpha_{0} \}.$ The transverse shift sector, including  $\{\bar{\beta}^{\perp},\bar{\Gamma}^{\perp},\bar{B}^{\perp}\}$, is also complete with characteristic speeds  $\{-\bar{\beta}^{s}_{0},-\bar{\beta}^{s}_{0} \pm \sqrt{g/\chi_{0}}\}$.
The light sector also has a complete set of eigenvectors including the projections $\{  \overline{\gamma}_{\perp\perp}, \bar{A}_{\perp\perp},\overline{\gamma}_{s\perp},\bar{A}_{s\perp}\}$, with characteristic speeds  $\{-\bar{\beta}^{s}_{0} \pm \alpha_{0}\}$. 

Finally, notice that the choice $g=3/4$ is especially delicate because the characteristic velocities of the longitudinal shift modes collapse to light speed: no complete set of eigenvectors can be found, and the strong pseudo-hyperbolicity of the system is spoiled. This might be a problem, at least around Minkowski spacetimes, as it has been reported previously by several authors~\cite{Z43,beyer,gundlach2}.

% % % % % % % % % % % % % % % % % % % % % % % % % %

\subsection{Klein-Gordon equation}

The evolution of a complex scalar field $\phi$ is described by the Klein-Gordon equation
\begin{eqnarray}\label{covariantKG}
  g^{ab} \nabla^a \nabla_b \phi = \frac{dV}{d|\phi|^2} \phi = V' \phi,
\end{eqnarray}
where $V(|\phi|^{2})$ is the potential depending only on the scalar field magnitude.
By using the 3+1 decomposition, and introducing $\Pi \equiv
-\mathcal{L}_{n} \phi$ as a new evolved field, the Klein-Gordon equation~(\ref{covariantKG}) can be written as
\begin{eqnarray}
  \partial_t \phi &=& \beta^k \partial_k \phi - \alpha \Pi,
\\
  \partial_t \Pi &=& \beta^k \partial_k \Pi \nonumber
   +  \alpha \left[ -\gamma^{ij} \nabla_i \nabla_j \phi + \Pi \,tr K   \right. \\\nonumber
   &+& \left.V' \phi \right]  - \gamma^{ij} \nabla_i \phi \nabla_j \alpha,
\end{eqnarray}
or, by using the conformal fields of the CCZ4 formalism,
\begin{eqnarray}
\partial_t \Pi &=& \beta^k \partial_k \Pi \nonumber
   +  \alpha \left[ -\chi \tilde{\gamma}^{ij} \partial_i \partial_j \phi + \chi \tilde{\Gamma}^k \partial_k \phi   \right. \\\nonumber
   & +& \left. \frac{1}{2} \tilde{\gamma}^{ij} \partial_i \phi \partial_j \chi  + \Pi\, tr K + V' \phi \right] 
\nonumber \\
&-& \chi \tilde{\gamma}^{ij} \partial_i \phi \partial_j \alpha \,\,.
\label{KG2}
\end{eqnarray}

In order to study self-gravitating boson stars we need to define the stress-energy tensor produced by this complex scalar field\footnote{Notice that there is a difference of a factor $1/2$ with respect to other definitions.}~\cite{mace}
\begin{eqnarray}
  T_{ab} &=& \nabla_{a} \phi^*\, \nabla_{b}\phi+\nabla_{b} \phi \, \nabla_{a} \phi^*  \\ \nonumber
& & - g_{ab}\left(g^{cd}\nabla_{c} \phi^*\, \nabla_{d} \phi + V\right),
\end{eqnarray}
where $\phi^*$ is the complex conjugate of $\phi$. One can also take advantage of the $U(1)$ symmetry to define the Noether charge, given by:
\begin{equation}
 N \equiv \int_{\Sigma_{t}} (-n_a J^{a}) \sqrt{\gamma}\,d^{3}x, 
\label{noe} 
\end{equation}
where
\begin{equation}
  J^{a} = i g^{ab} (\phi^*\,\nabla_{b}\phi - \phi\,\nabla_{b} \phi^*) .%\\
%   &= 2(  \Pi_{R} \phi_I - \Pi_I \phi_R )
%  \nonumber
\end{equation}
The Noether charge can be interpreted as the number of bosonic particles~\cite{liebpa}.

% % % % % % % % % % % % % % % % % % % %

\section{Numerical tests }\label{nts}

Our modification of the CCZ4 formalisms is examined by evolving known solutions --robust stability, gauge waves and single boson star--, some of which are included as standard testbeds~\cite{apples}. Here we describe first the setup of our simulations and present the tests results.

\subsection{Setup }
We adopt finite difference schemes, based on the Method of Lines, on a regular Cartesian grid. A fourth order accurate spatial discretization --satisfying the summation by parts rule--, together with a third order accurate (Runge-Kutta) time integrator, are used to achieve stability of the numerical implementation~\cite{ander}.
To ensure  sufficient resolution within the compact objects (i.e., boson stars in this work) in an efficient manner, we employ adaptive mesh refinement (AMR) via the \textsc{had} computational infrastructure that provides distributed, Berger-Oliger style AMR \cite{had,lieb} with full sub-cycling in time, together with an improved treatment of artificial boundaries \cite{lehner}. 
We adopt a Courant parameter of $\lambda_c \approx 0.25$ such that $\Delta t_{l} = \lambda_c \, \Delta x_{l}$ on each refinement level $l$ to guarantee that the Courant-Friedrichs-Levy condition is satisfied.
This code has been used extensively for a number of other projects and it has already been rigorously tested.

%In our implementations (\MB{or code}) we have added a new parameter, which we are going to call \textit{enforce}, if we activate this parameter we are be able to impose the conformal constraint \eqref{enf} at every integration step like \cite{alcub3} in BSSN code.

% % % % % % % % % % % % % % % % % % %
\subsection{Robust stability test}
% % % % % % % % % % % % % % % % % % %

\begin{figure}
\centering
\includegraphics[width=1.0\linewidth]{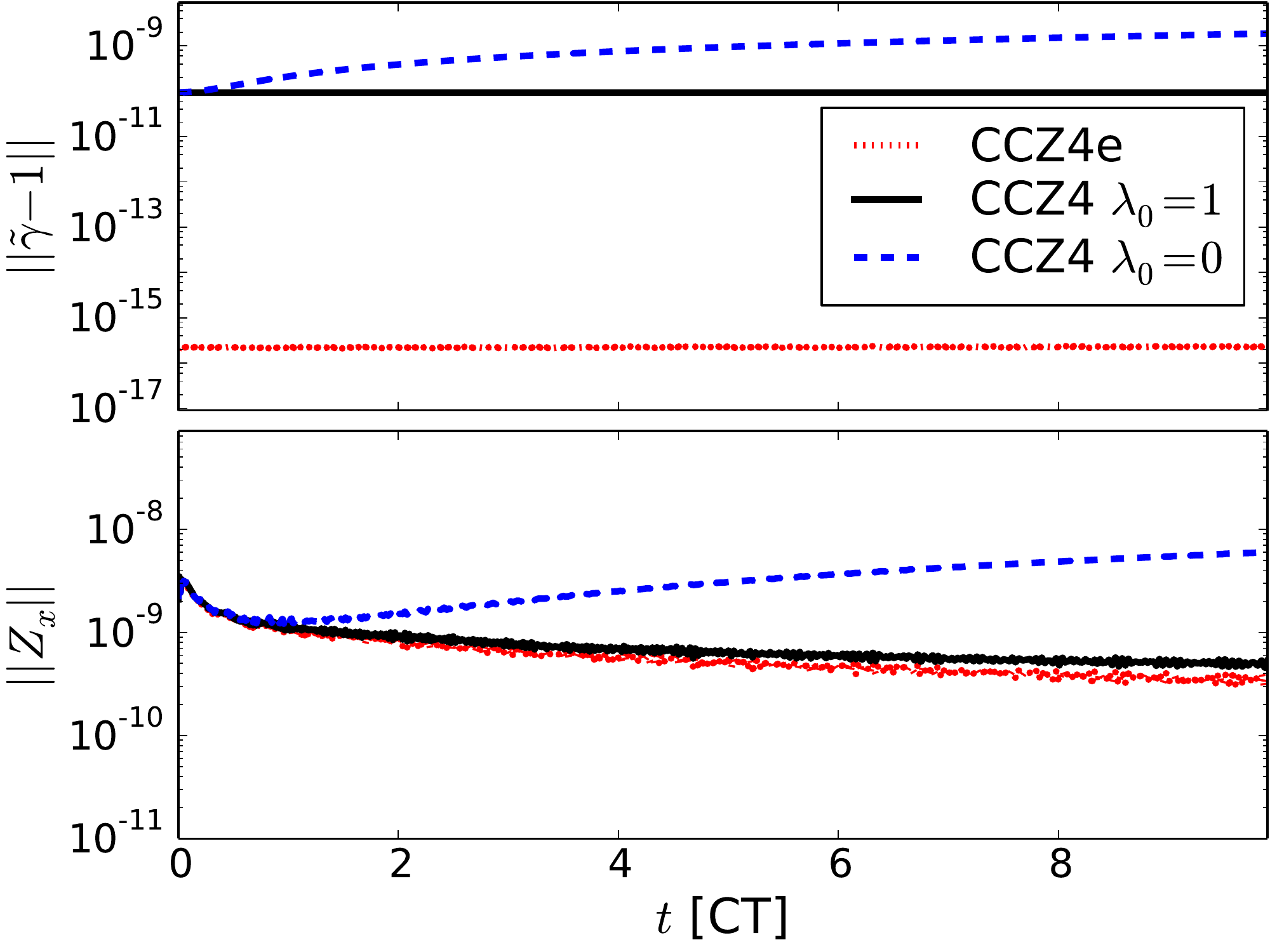}
\caption{{\em Robust stability}. $L_{2}$-norm of $|\tilde{\gamma}-1|$ (top panel) and $|Z_{x}|$ (bottom panel) as a function of time -in crossing time units-. 
Some modes increase for the CCZ4 system with $\lambda_{0}=0$ (blue dashed line), showing that this choice leads to a  weakly pseudo-hyperbolic system. These modes (and all others) remain constant --a sign of the strong hyperbolicity of the system-- in the other two cases; CCZ4 with $\lambda_{0}=1$ (black solid line) and CCZ4e (red dotted line), where the conformal constraints are algebraically enforced.}
\label{l2norm_CCZ4}
\end{figure}

We first carry out the robust stability test-bed~\cite{alcub2,CCC} to confirm the characteristic structure of our system  (\ref{syseq1}-\ref{syseq2}). The test consists on a Minkowski background metric plus a small random perturbation in each of the evolution fields, such that only the principal part and the linear terms are significant (i.e., the scalar field potential $V$ and all the damping coefficients are set to zero in this test). A linear growth on any field indicates a weakly hyperbolic system. We set a 2D domain $[-0.5,0.5]^2$ with periodic boundaries conditions and $N=100$  grid points in each direction. No artificial  Kreiss-Oliguer dissipation is included for this test. As it was shown in subsection~\ref{hyperCCZ4}, the hyperbolicity of the system depends on
the parameter $\lambda_{0}$, so we analyze the effect of this parameter. Besides, we use $f = 2/\alpha$, $g=3/4$ and $\eta= 2$, that is, the $1+\log$ slice with common values for the Gamma-driver shift condition.

The $L_{2}$-norms of some constraints are displayed in Fig.~\ref{l2norm_CCZ4} for three cases. In the first one (CCZ4e) the conformal constraints are algebraically enforced after each timestep, as it is currently done in all the flavors of BSSN and conformal Z4. The other two cases (CCZ4) corresponds to $\lambda_{0}=0$ and $\lambda_{0}=1$ without any algebraic enforcing. Our simulations show a linear growth on  $||\tilde{\gamma}-1||$ that propagates to $||Z_{x}||$ and, eventually, to all the other fields. This linear growth indicates a lack of strong hyperbolicity of the system for $\lambda_{0}=0$. All the norms are constant both for CCZ4e and for CCZ4 with $\lambda_{0}=1$, as it is  expected for a well-posed system. Henceforth, we are going to use the choice $\lambda_{0}=1$ for all the forthcoming simulations with CCZ4.

% % % % % % % % % % % % % % % % % % %
\subsection{Gauge waves}
% % % % % % % % % % % % % % % % % % %

A family of non-trivial exact solutions can be constructed by performing a coordinate transformation on $\{x,t\}$ to  Minkowski spacetime. The resulting line element can be written as~\cite{alcub2,CCC,alic,apples}
\begin{equation}
  ds^2 = - H(x-t)dt^2 + H(x-t) dx^2 + dy^2 + dz^2,
  \nonumber
\end{equation}
where $H(x-t) = 1 - A \sin \left[ k (x -t) \right]$.
We set an amplitude $A=0.1$ and a wave number $k = 2 \pi/L$, being $L$ the size of the domain. This solution is exact with harmonic slicing and zero shift, corresponding to the choice $f=1$ and $g=0$. The domain for this one-dimensional test is $[0,1]$ with periodic boundary conditions and $100$ grid points.

\begin{figure}
\centering
\includegraphics[width=1.0\linewidth]{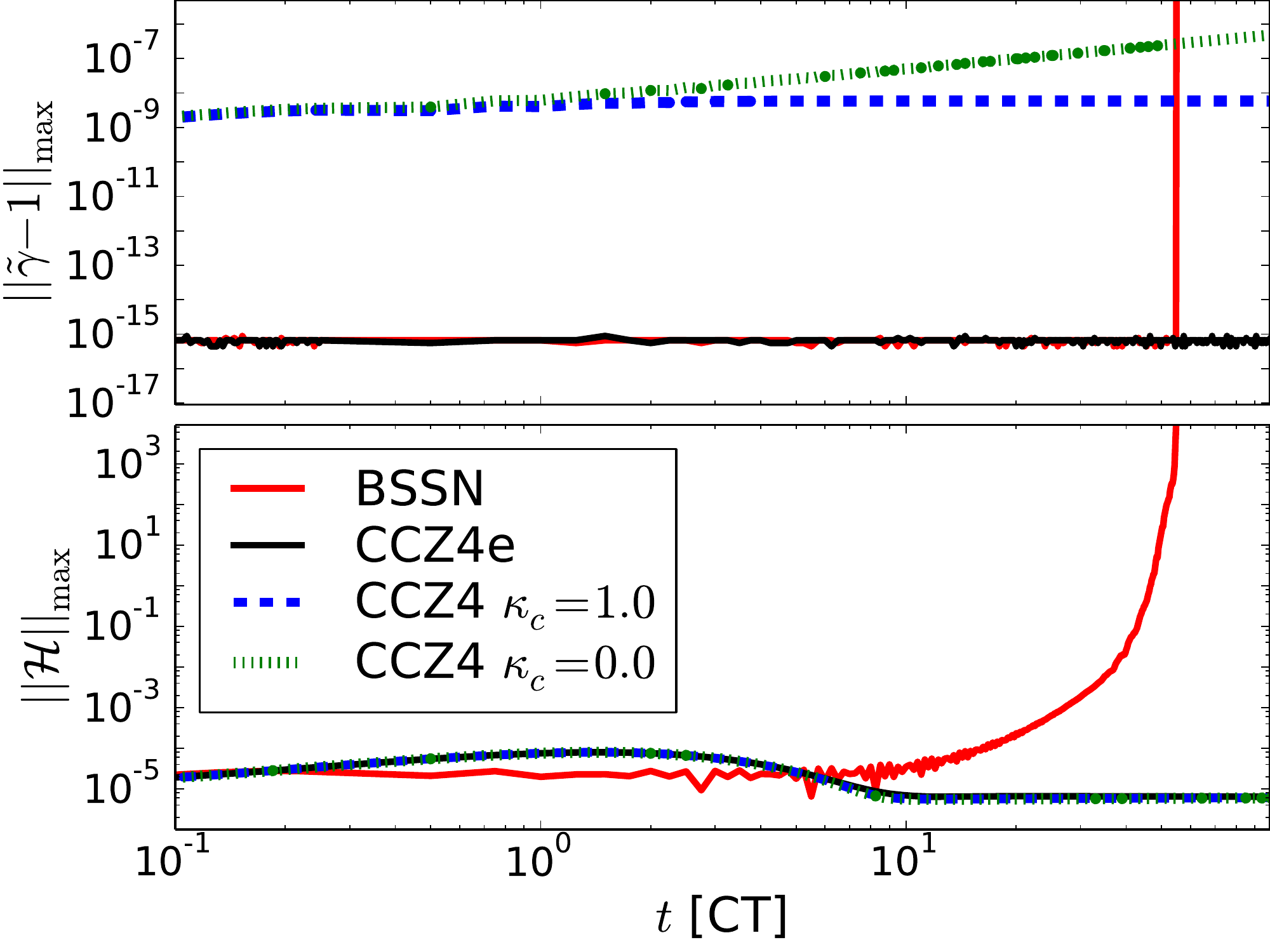}
\caption{   {\em Gauge waves}. $L_{\infty}$-norm of $|\tilde{\gamma}-1|$ (top panel) and $|\mathcal{H}|$ (bottom panel) as a function of time -in crossing time units -. The BSSN system (red solid line) and CCZ4 with $\kappa_{c}=0$ (green dotted line) display an unbound growth in some constraints. Both CCZ4e (black solid line) and CCZ4 with $\kappa_{c}=1$ (blue dashed line) maintain the constraints under control at least for $100$ crossing times.}
\label{max_norm_ham}
\end{figure}

$L_{\infty}$-norm for some constraints are shown in Fig.~\ref{max_norm_ham} for four different cases, which can be compared with figure 1 in~\cite{alic}.
In the first two cases (i.e., BSSN and CCZ4e), the conformal constraints are algebraically enforced after each time step. The other two cases correspond to CCZ4 with  either $\kappa_{c}=1/L$ or $\kappa_{c}=0$ (in both cases $\kappa_{z}=1/L$). It is clear that the BSSN formulation fails this test, as both the conformal and the Hamiltonian constraint suffer exponential growth. In contrast, all the constraints remain under control when using the CCZ4e. The most important outcome of this test is the fact that the CCZ4 formulation with $\kappa_{c}=1$ is also stable, meaning that it is not required to enforce algebraically the conformal constraints to keep them under control. The last case, CCZ4 with $\kappa_{c}=0$, presents a linear growth in the conformal constraint $|\tilde{\gamma} - 1|$, which unavoidably will lead to a failure due to the propagation to other fields. The same behavior is observed in simulations on generic spacetimes, indicating that the choice of the damping coefficients  $\{\kappa_{z},\kappa_{c}\}$ is crucial to achieve accurate and stable solutions.

% % % % % % % % % % % % % % % % % % %
\subsection{Single solitonic boson star}\label{sbs}
% % % % % % % % % % % % % % % % % % %

The initial data for complex scalar field configurations in spherical symmetry can be solved numerically for the static metric~\cite{liebpa,mace}
\begin{equation}
   ds^2 = -\alpha^{2}(r) dt^{2} + \psi^{4}(r) (dr^{2} + r^{2} d\Omega^2),
\label{metric_bs}
\end{equation}
by adopting the following harmonic ansatz for the scalar field
\begin{equation}
   \phi(t,r) = \phi_{0}(r)\,e^{-i\omega t} .
\label{scf}
\end{equation}  

\begin{figure} 
\centering
\includegraphics[width=1.0\linewidth]{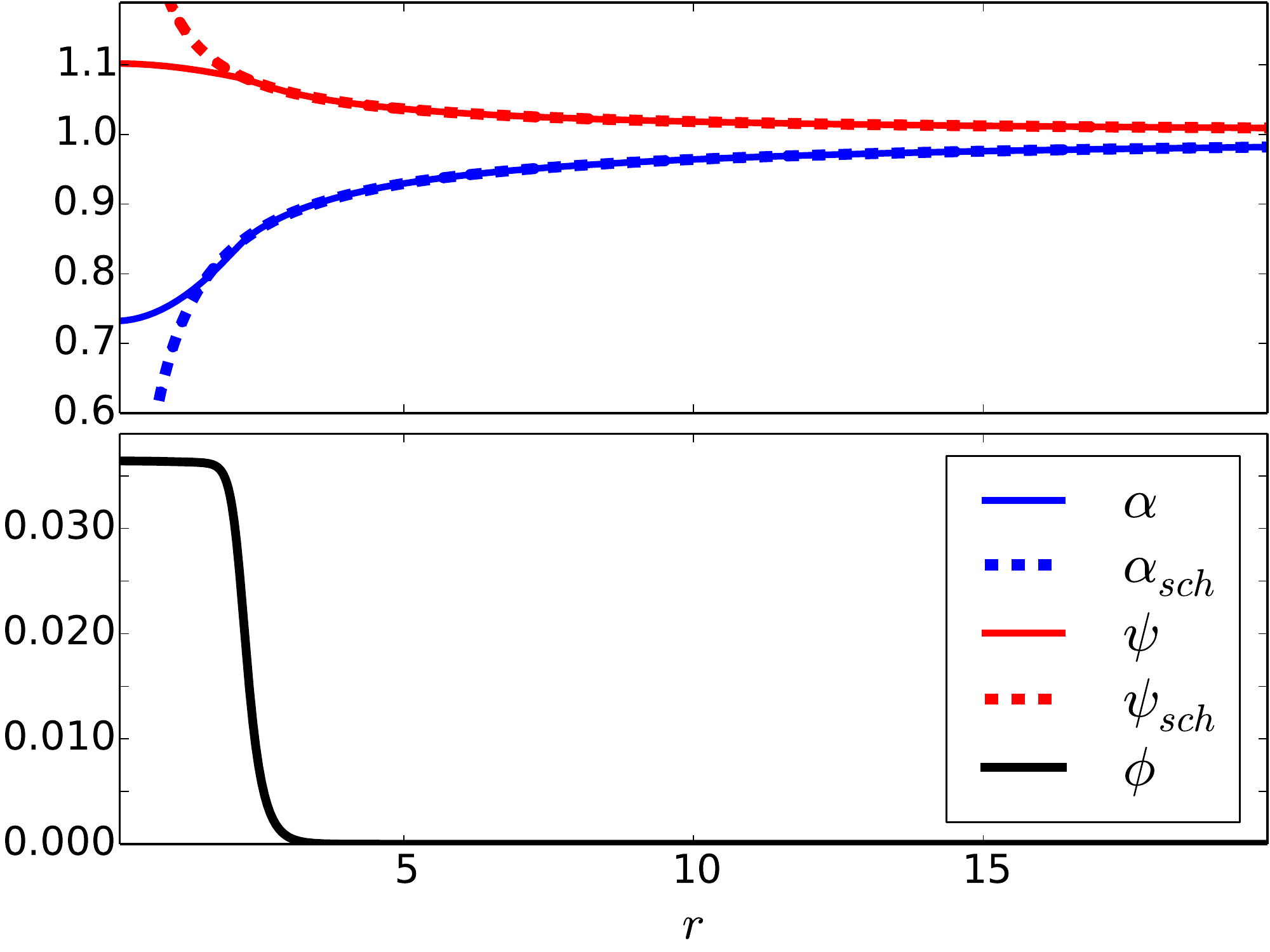}
\caption{ {\em Solitonic boson star}. The top panel displays the metric components $\alpha(r)$ (blue solid line) and $\psi(r)$ (red solid line) for the typical solitonic boson stars with compactness $C\approx 0.118$ used here, compared to the Schwarzschild solution (dashed lines) in isotropic coordinates. The bottom panel shows the scalar field profile $\phi_{0}(r)$, which is almost constant in the interior and decays rapidly at the surface of the star.}
\label{phi_alp_psi}
\end{figure}

Within these assumptions the EKG system reduces to a set of ordinary differential equations (ODEs) that can be solved by imposing appropriate boundary conditions (i.e., regularity at the origin and asymptotically flat at large distances). The equations are further simplified by using polar-areal coordinates. Therefore, the standard procedure is to solve the equations in these coordinates and then perform a (numerical) coordinate transformation into isotropic coordinates, which can be transformed easily to Cartesian ones~\cite{2004PhDT.......230L}. 

\begin{figure}
\centering
\includegraphics[width=1.0\linewidth]{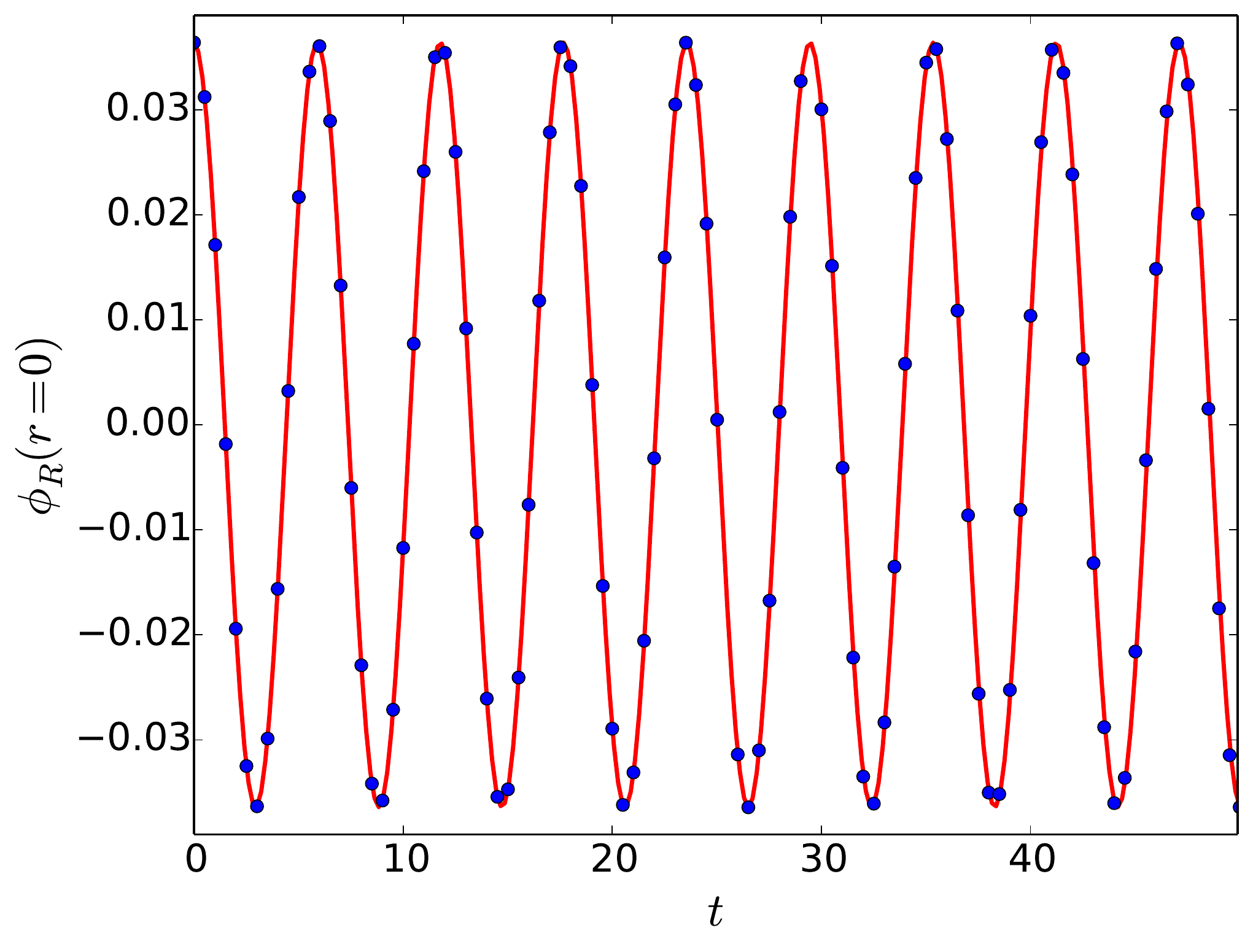}
\caption{ {\em Solitonic boson star.} Evolution of the real part of $\phi$ at $r=0$. The solid red line illustrate the analytically expected value $\cos(\omega t)$ with $\omega=1.0666.$ The blue circles show the numerically solution obtained with different evolution systems (i.e., BSSN and CCZ4), which can not be distinguished by eye in this plot.}
\label{phiR}
\end{figure}

Different interaction potentials $V(|\phi|^{2})$ lead to boson stars with different compactness~\cite{liebpa}. We are interested in a particular family of very compact boson stars, commonly known as nontopological solitonic boson stars~\cite{frie,1992PhR...221..251L}, where the potential is given by 
\begin{equation}
V(|\phi|^{2}) = \mu^{2} |\phi|^{2}\left(1 - 2\,\frac{|\phi|^{2}}{\sigma_{0}^{2}}\right)^{2} \,.
\end{equation}
Here $\sigma_{0}$ is a constant that determines the compactness of the star and $\mu$ is related to the scalar field mass. By setting $\sigma_{0}=0.05$ and the scaling factor $\mu \sigma_{0}\sqrt{8\pi}=1$,  a suitable stable equilibrium configuration can be obtained for $\phi_0(r=0)=0.0364$ and $\omega=1.0666$. The resulting star has mass $M=0.36$ and radius $R=3.08$, so its compactness is $C=M/R=0.118$. This configuration is well inside the stable branch, since the most massive stable star has a mass $M_{max}\approx 1.84$.
The profiles of $\alpha(r)$, $\psi(r)$ and $\phi_0(r)$ for this particular solution are plotted in Fig.~\ref{phi_alp_psi}.

This configuration is evolved in a domain $[-16,16]^{3}$ with radiative boundary conditions. There are $60$ grid points in each direction and three refinement levels, such that the highest resolution is $\Delta x=0.1$. The simulations are performed by using the CCZ4 formulation with $\kappa_{z}=0.1$ and either $\kappa_{c}=1$ or $\kappa_{c}=0$. We also include the solutions obtained with the BSSN formulation for comparison purposes.

\begin{figure}
\centering
\includegraphics[width=1.0\linewidth]{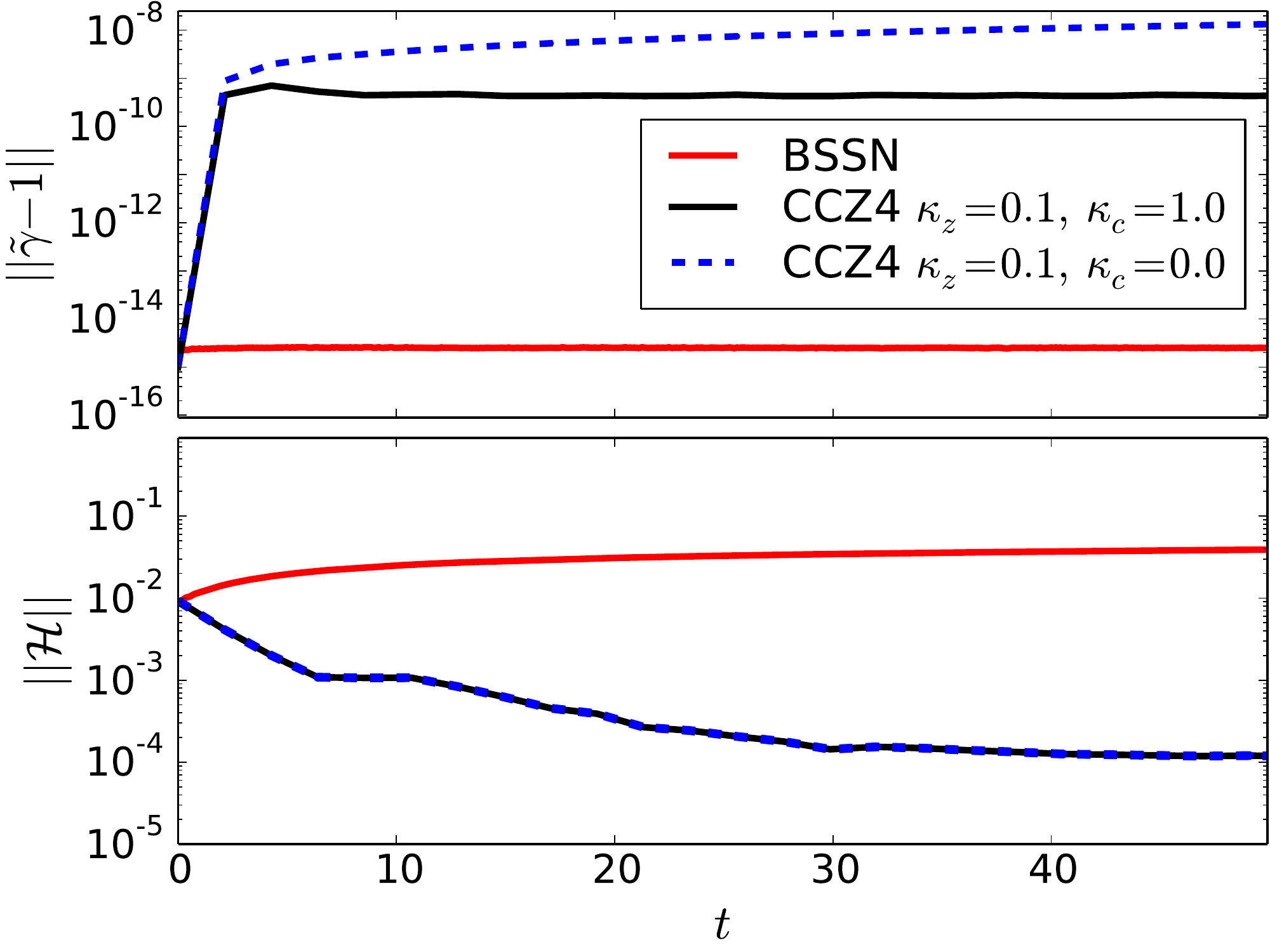}
\caption{ {\em Solitonic boson star.}  $L_{2}$-norm of $|\tilde{\gamma}-1|$ (top panel) and $|\mathcal{H}|$ (bottom panel) as a function of time. The solution obtained with BSSN (red solid line) shows a small $|\tilde{\gamma}-1|$ constraint as a result of enforcing constraint in each integration time-step, but the Hamiltionian constraint increase over time. The solutions obtained with CCZ4 are stable if we add the damping terms for the conformal constraints (black solid line) --otherwise there is a linear growth in $|\tilde{\gamma}-1|$ that will lead to a unstable evolution (blue dashed line).}
\label{ham_det_bs}
\end{figure}

The evolution of the scalar field real part $\phi_R(t,r=0)$ is displayed in Fig.~\ref{phiR}, together with the expected analytical behavior $\phi_{0}(r_{0})\cos(\omega t)$. The solutions for all the cases considered, either with BSSN or CCZ4, show a very good agreement with the analytical expectation. Differences arise however in the $L_{2}$-norm of some constraints, plotted in Fig.~\ref{ham_det_bs}. Both the conformal and the physical constraint remain under control by using either BSSN or CCZ4 with $\kappa_{z}=0.1$ and $\kappa_{c}=1$. However, notice that the errors of the physical constraints obtained with the CCZ4 and this parameter choice are several orders of magnitude smaller than the ones obtained by using BSSN.

%\CP{Finally, we include a convergence plot }

% % % % % % % % % % % % % % % % % % % % % % % % % % % % 
% % % % % % % % % % % % % % % % % % % % % % % % % % % % 
\section{Dynamics of solitonic boson stars}
\label{dsbs}
% % % % % % % % % % % % % % % % % % % % % % % % % % % %
% % % % % % % % % % % % % % % % % % % % % % % % % % % %

Here the dynamics of binary boson stars is studied, focusing on the final state of the system after the merger. We first describe how to construct initial data for a binary boson star system with generic angular momentum. We consider several head-on cases, and finish with orbiting binary systems. 

\subsection{Initial data and setup}

We extend the procedure describe in~\cite{shi} to construct accurate boosted initial data from an spherically symmetric solution. Our starting point is the line element given by 
\begin{eqnarray}
ds^2 = -\alpha_{0}^{2}dt_{0}^{2} + \psi_{0}^{4}(dx_{0}^{2}+dy_{0}^{2}+dz_{0}^{2}),
\end{eqnarray}
where $\alpha_{0}=\alpha_{0}(r_{0})$ and $\psi_{0}=\psi_{0}(r_{0})$, being $r_{0}=\sqrt{x_{0}^{2}+y_{0}^{2}+z_{0}^{2}}.$ By performing a Lorentz transformation $t=\Gamma(t_{0} + vx_{0}), \,x=\Gamma(x_{0} + vt_{0})$ (being $\Gamma=1/\sqrt{1-v^2}$), one can obtain 
\begin{eqnarray}
ds^2 & = &  -\Gamma^{2}(\alpha_{0}^{2} - \psi_{0}^{4}v^{2})dt^{2} + 2\Gamma^{2}\,v\,(\alpha_{0}^{2} - \psi_{0}^{4})dtdx \nonumber\\
& &+ \psi_{0}^{4}(B_{0}^{2}dx^{2}+dy^{2}+dz^{2}) ~~,
\end{eqnarray}  
The gauge choice is given by
\begin{eqnarray}
\alpha = \frac{\alpha_{0}}{B_{0}} \,\, , \,\,
\beta^{x} &=& \left(\frac{\alpha_{0}^{2} - \psi_{0}^{4}}{\psi_{0}^{4} - \alpha_{0}^{2}v^{2}  }\right)v ~~,
\end{eqnarray}
with $B_{0}=\Gamma\,\sqrt{\left(1-\frac{v^2\alpha_{0}^{2}}{\psi_{0}^{4}}\right)}$.
Notice that $r_0$ can be written in terms of the new coordinates, namely $r_{0}=\sqrt{\Gamma^{2}(x-vt)^{2}+y^{2}+z^{2}}$. 

Now we only have to perform the Lorentz transformation to the scalar field quantities. First, the harmonic ansatz given by eq.~\eqref{scf} can be generalized, to allow for non-identical boson stars, by including a phase shift $\theta$ and the direction of rotation  $\epsilon=\pm 1$\cite{liebpa}, namely:
\begin{eqnarray}
  \phi(t_{0},r_{0}) &=& \phi_{0}(r_{0})\,e^{-i(\epsilon\omega t_{0} + \theta)} \label{azshift}
\end{eqnarray}

We can compute the field $\Pi(t,r_0)$ from its definition, calculated in the boosted frame. The final expressions, evaluated at $t=0$, are
\begin{eqnarray}
  &\phi_{R}&(r_{0}) =\phi _0\left(r_0\right) \cos (\theta-\Gamma  v x \epsilon\omega  ),\\
  &\phi_{I}&(r_{0}) =-\,\phi _0\left(r_0\right) \sin (\theta-\Gamma  v x \epsilon\omega ),\\
  &\Pi_{R}&(r_{0}) =\frac{ \Gamma \, \epsilon\,\omega  \phi _0\left(r_0\right) (\beta^{x}  v+1) \sin (\theta - \Gamma  v x \epsilon\omega) }{\alpha} \nonumber\\ 
  & +&\frac{  \Gamma^2 x (\beta^{x} +v) \phi _0'\left(r_0\right) \cos (\theta - \Gamma  v x \epsilon\omega )}{\alpha  r_0},\\
  &\Pi_{I}&(r_{0}) = \frac{\Gamma \, \epsilon\,\omega  \phi _0\left(r_0\right) (\beta^{x}  v+1) \cos (\theta - \Gamma  v x \epsilon\omega) }{\alpha}\nonumber\\ 
  &-&\frac{ \Gamma^2 x (\beta^{x} +v) \phi _0'\left(r_0\right) \sin (\theta - \Gamma  v x \epsilon\omega )}{\alpha  r_0} ~~.
\end{eqnarray}

We are mainly interested on binary systems. Along the lines described  in~\cite{pale1}, initial data for head-on binaries can be constructed as the superposition of two single solitonic boson star solutions, located at positions $r_{i}$, in the following way
\begin{align}
\phi_0(r) &= \phi^{(1)}_{0}(r_{1})e^{-i\omega t} +  \phi^{(2)}_{0}(r_{2})e^{-i(\epsilon\omega t + \theta) }
\\
\alpha(r)&= \alpha^{(1)}(r_{1}) +  \alpha^{(2)}(r_{2}) -1 \\
\psi(r)&= \psi^{(1)}(r_{1}) +  \psi^{(2)}(r_{2}) -1      
\end{align}
where superindex $(i)$ indicate each star.
The initial data for orbiting binaries can be constructed in an analogous way, but using the boosted boson star --with velocity $v$ along the $x$-direction-- as the basic solution and performing the superposition of all the non-trivial evolution fields.

For the head-on cases each solitonic BS is centered at $(\pm x_c,0,0)$, while that for the orbiting case they are centered at $(0,\pm y_c,0)$, being $x_c=y_c=4$. In both cases the cubical domain is given by $[-60,60]^{3}$ and $120$ grid points in each axis, leading to a coarsest resolution $\Delta x_{0} = 1.0$. We set five refinement levels such that the last one, covering both stars, has a resolution $\Delta x_{4}=0.0625$ (i.e., there are approximately 96 points covering each star). 

Besides the constraints, additional analysis quantities are evaluated in our binary boson stars simulations: the Noether charge or boson number, given by the volume integral eq.~(\ref{noe}), and the ADM mass and angular momentum as described in the Appendix. The  surface integrals, eq.~(\ref{admmass},\ref{admJz}), are computed in a sphere located at $R_{ext}=40$.  

% % % % % % % % % % % % % % % % % % % % % % % % % %
\subsection{Head-on cases}
% % % % % % % % % % % % % % % % % % % % % % % % % %

We consider initial data constructed as described in the previous subsection with the ansatz~\eqref{azshift} and no boost. Even for a particular solution (i.e., the one described in subsection~\ref{sbs}), located at fixed initial positions, there is an infinite family of configurations depending on the parameters $\{\theta, \epsilon\}$. We differentiate two essentially different scenarios: (i) two identical boson stars, that is, with the same phase direction of rotation (i.e., $\epsilon=+1$), that will be denoted as B-B($\theta$), and (ii) a binary formed by a boson and an anti-boson star, denoted as B-aB($\theta$), such that their phases have opposite rotation direction (i.e., $\epsilon=- 1$) and their Noether charges have opposite sign. We consider four different cases for B-B($\theta$) with phase shifts $\theta=\{0,\pi/2,\pi,3\pi/2\}$, and two for B-aB($\theta$) with $\theta=\{0,\pi\}$. 

\begin{figure}
\centering
\includegraphics[width=7.8cm,height=7.0cm]{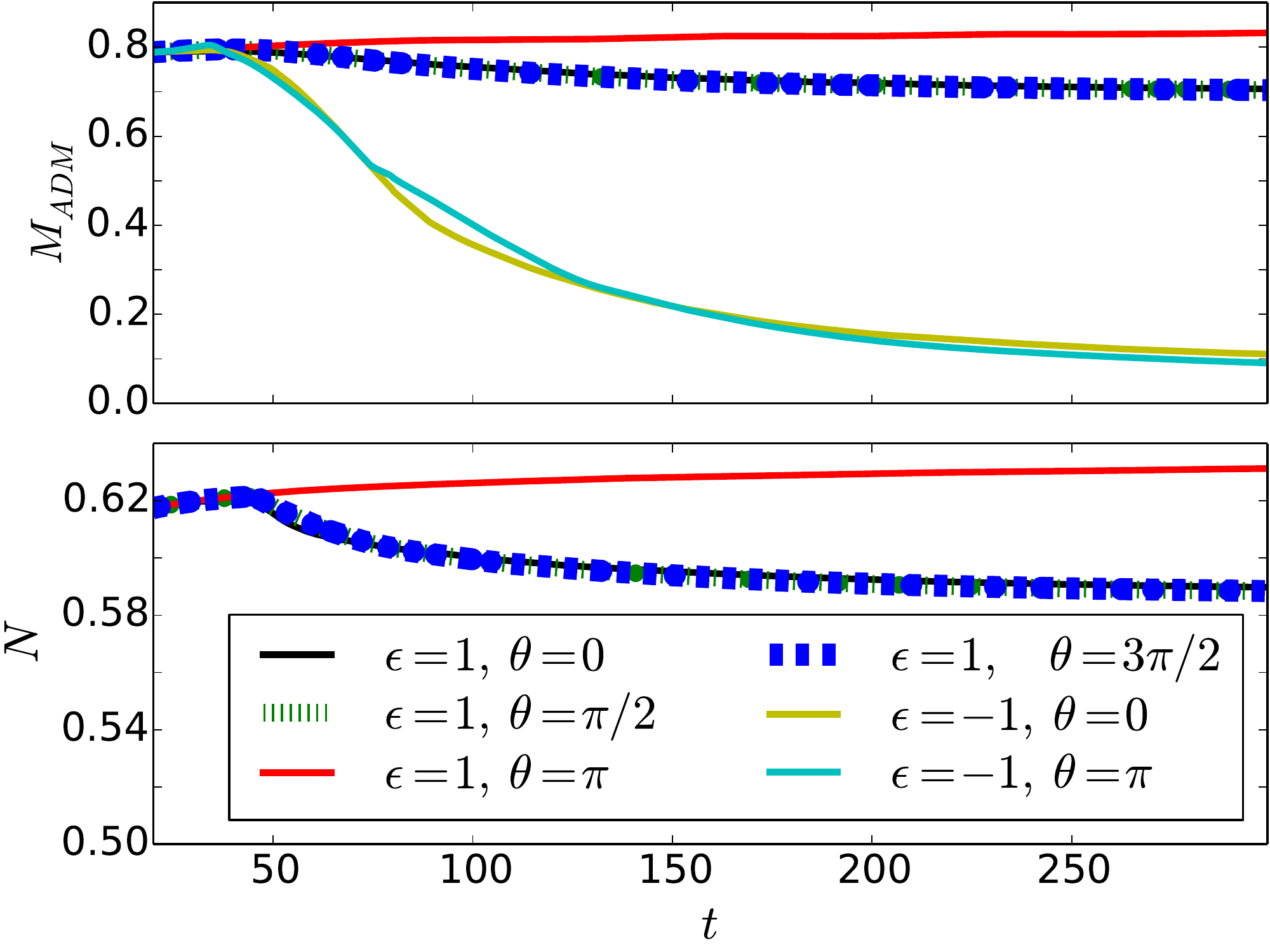}
\caption{{\em Head-on binary collisions.} ADM mass and Noether charge as a function on time for the different cases studied. The boson-boson binaries merging into a single one losses approximately a $5\%$ of their initial mass and Noether charge. In contrast, the boson-antiboson binaries annihilate during the merger, radiating most of the scalar field (and the corresponding mass). The total Noether charge for the B-aB cases is zero through all the evolution. }
\label{headonmassnoe}
\end{figure}

\begin{figure*}
\centering
\includegraphics[width=15cm, height=20cm]{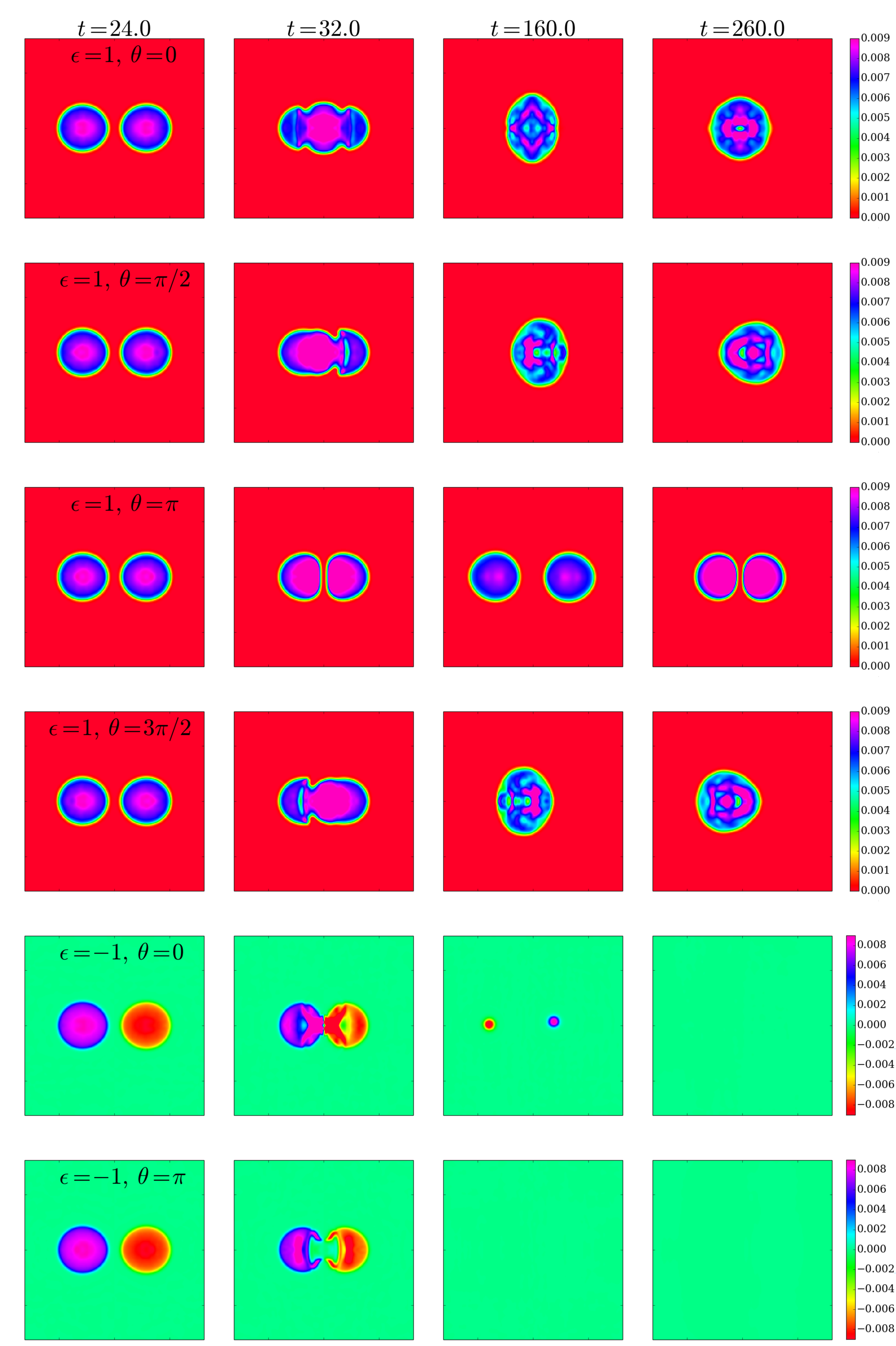}
\caption{{\em Head-on binary collisions.} Time snapshots of the Noether charge in the plane $z=0$. Each row corresponds to the different B-B($\theta$) and B-aB($\theta$) cases studied here. The collision of the stars happens approximately at $t=28$. The result of the B-B is a single boson star except in the case of B-B($\pi$). The stars in the B-aB case annihilate each other during the merger.}
\label{hn_final}
\end{figure*}

The evolution of the ADM mass and the Noether charge is presented in Fig.~\ref{headonmassnoe}, and some snapshots of the evolution for all cases are displayed in Figure~\ref{hn_final}. Let us start by describing the dynamics of the binaries formed by two boson stars. The collision of the B-B($0$), B-B($\pi/2$) and B-B($3\pi/2$) cases leads to a merger, resulting into a single solitonic boson star with roughly the same total initial mass and  Noether charge (i.e., except a small fraction that is emitted by gravitational waves and scalar field radiation, see Fig.~\ref{headonmassnoe}). However, the scalar field interaction in the B-B($\pi$) produces a repulsive force that overcomes the gravitational attraction. Therefore, the binary suffers several inelastic collisions before relaxing to a system with two touching stars --which do not merge into a single one.
The binaries formed by a boson and an antiboson star annihilates each other during the merger for the two extreme phase shifts considered (i.e., $\theta=0,\pi$). Most of the scalar field is radiated away to infinity, and only a small fraction remains near the region of the collision.  % A summary of the parameters and the results for this cases can be found in  Figure~\ref{headonmassnoe}.

%\begin{table*}
%\label{bs}
%\begin{ruledtabular}
%\begin{tabular}{ccccccc}
%  $\epsilon$ & $\theta$ & $N_{t=0}$ & $N_{t=300}$ & ${M_{\text{ADM}}}_{t=300}$ 
%\\ \hline\hline 
% 1 & 0 & 0.61952305285  & 0.58973609323 &   0.709469639072   \\ 
% 1 & $\pi/2$  & 0.61952304463  & 0.58838837716 & 0.706371652112   \\ 
% 1 & $\pi$  & 0.61952303641  & 0.63125103112 & 0.829449184423   \\ 
% 1 & $3\pi/2$  & 0.61952304463  & 0.58841972176 &  0.706133002915   \\ 
%  -1 & 0  & -0.29629076970e-14 & -0.53396985810e-04   &  0.123305868601  \\
%  -1 & $\pi$  & -0.21094237468e-14  & -0.12450197646e-03 & 0.104613354165 
%\end{tabular}
%\caption{{\em Results of Hean-on cases evolution.} The table is shown: direction of rotation, phase shift, initial Noether charge and Noether charge and ADM mass (at $r=40$ extraction surfaces) in the las time evolution. The initial ADM mass was 0.75862650929 it was the same for all configurations, we can observer  in the cases when there  were merger we lost approximately a $4\%-6\%$ mass, in contrast with the B-aBS($\theta$) case where final ADM mass was $88\%$ lower approximately.  The Noether charge shown that our binary simulation when merger happened has almost kept  constant  through the time.}
%\end{ruledtabular}
%\end{table*}

% % % % % % % % % % % % % % % % % % % % % % % % % %
\subsection{Orbiting cases}
% % % % % % % % % % % % % % % % % % % % % % % % % %

The initial data for the orbiting cases is constructed as described in subsection~\ref{dsbs} for identical stars (i.e., with $\epsilon=1,\theta=0$). We have considered different Lorentz boost velocities $v=\{0,0.05,0.10,0.15\}$ along the $x$-direction.
The last case corresponds to a binary almost in quasi-circular orbit. 

\begin{figure}
\centering
\includegraphics[width=8.2cm,height=7.0cm]{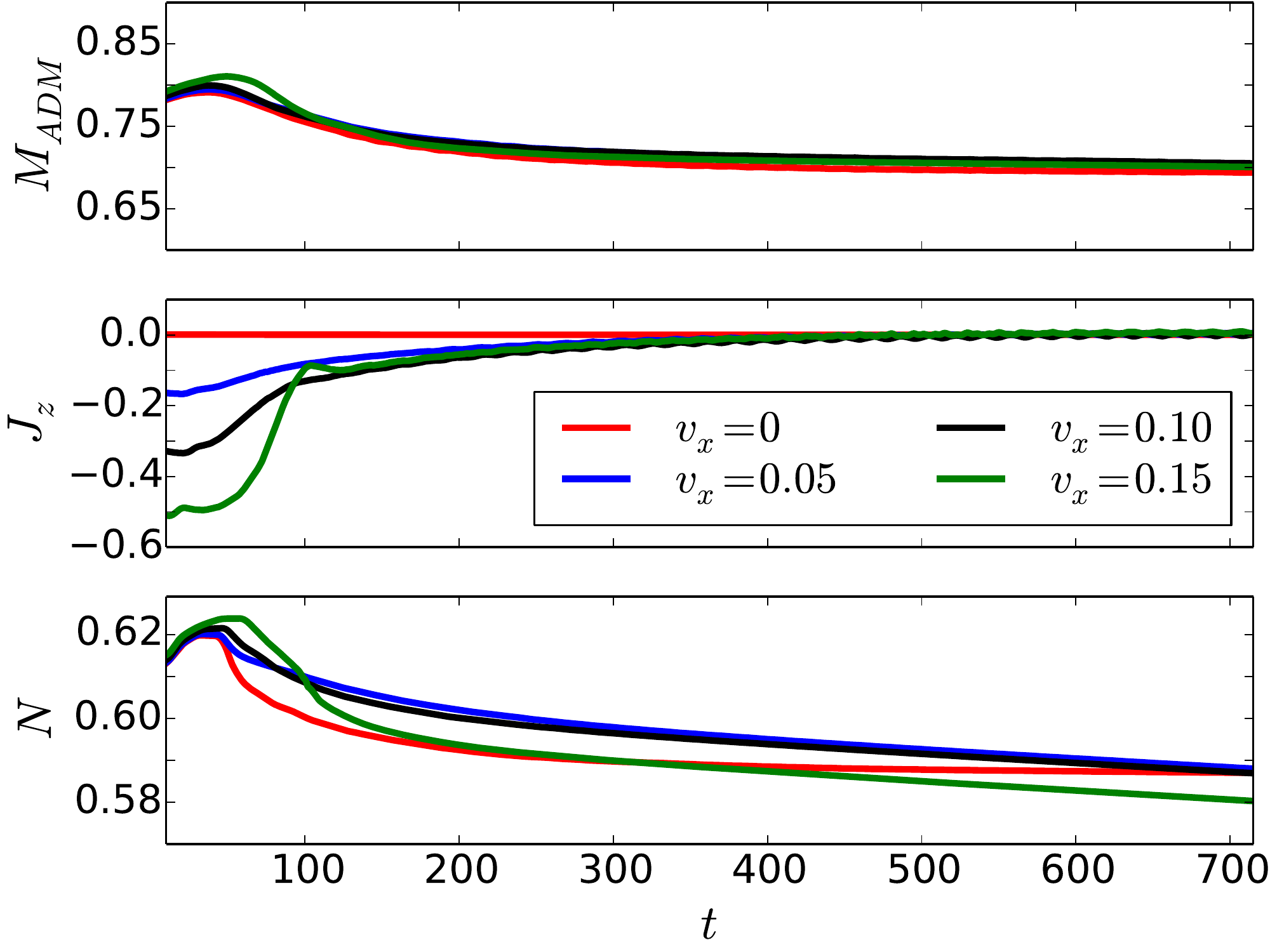}
\caption{{\em Orbital binary collisions.} ADM mass (top pannel), angular momentum $J_{z}$ (middle panel) and Noether charge (bottom panel) as a function on time for the different tangential boost velocities. During the coalescence approximately $5\%$ of the mass and Noether charge is lost, and almost all the angular momentum.}
\label{analysisquant}
\end{figure}

\begin{figure*}
\centering
\includegraphics[width=15cm, height=15cm]{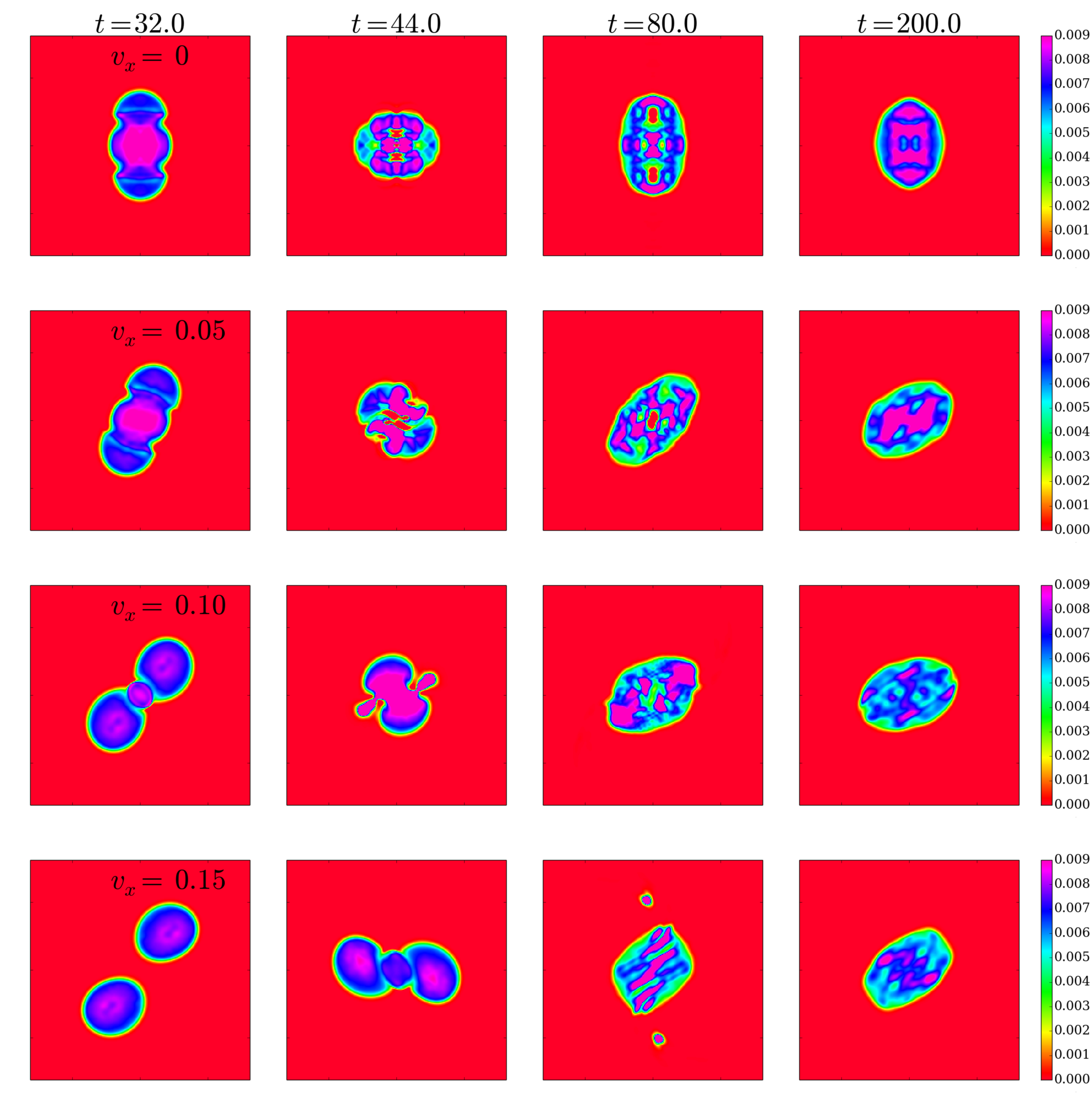}
\caption{{\em Orbital binary collisions.} Noether charge in the plane $z=0$ for the different boost velocities $v=\{0,0.05,0.10,0.15\}$. The merger between solitonic boson stars happens approximately at $t\approx 30$ for the cases $v=\{0,0.05,0.10\}$ and at $t\approx 40$ for the quasi-circular orbit case $v=0.15$. For the latter case, after the merger two blobs of scalar field take away a large fraction of the angular momentum from the system.}
\label{boost_final}
\end{figure*}

The ADM mass, the Noether charge and the angular momentum are shown in Fig.~\ref{analysisquant}, whereas few snapshots of the evolutions are displayed in Figure~\ref{boost_final}. All the cases --with angular momentum-- merge and form a rotating bar that quickly losses angular momentum and settles down to a non-rotating boson star. 

Although the system only losses a small fraction of the mass and the boson number during the coalescence, all the angular momentum is emitted by gravitational waves and scalar field radiation soon after the merger. This prompt radiation of angular momentum is most extreme for the system in quasi-circular orbit, corresponding to $v=0.15$: after the merger, two blobs of scalar field are ejected from the system at almost light speed, carrying a small fraction of the mass and boson number but a large amount of angular momentum (i.e., see $t=80$ of the $v=0.15$ in Figure~\ref{boost_final}, and the sudden drop of angular momentum in Fig.~\ref{analysisquant}).

This behavior can possibly be explained by the quantization of (adimensional) angular momentum $J^{rot}_z = k N$ (being $k=$ an integer) for rotating boson star configurations, which might be difficult to achieve exactly by a dynamical merger.  The resulting star from the merger should have at least $J^{rot}_z(k=1) \approx 0.62$ in order to correspond to a stable rotating boson star (see the bottom panel of Fig.~\ref{analysisquant} at $t\approx 50$). Although the (adimensional) angular momentum of the system $J_z/M^2 \approx 0.78$ is larger than $J^{rot}_z(k=1)$, the system does not relax to that state but to the one with the lowest angular momentum $k=0$. Stated in a different way, the system does not dynamically evolve towards a rotating boson star configuration, but towards a non-rotating one.

Although this seems in contradiction with the results in~\cite{pale2}, where a rotating boson star seemed to be produced, there are two important differences. First, these stars are much more compact, so the dynamics might be more dominated by non-linear gravitational effects. Second, due also to the high compactness, the dynamics is faster (i.e., the crossing time is shorter), so we can follow the evolution until a stationary state is achieved, which might not had been possible with the mini-boson stars.

\begin{figure}
\centering
\includegraphics[width=7.8cm,height=5.0cm]{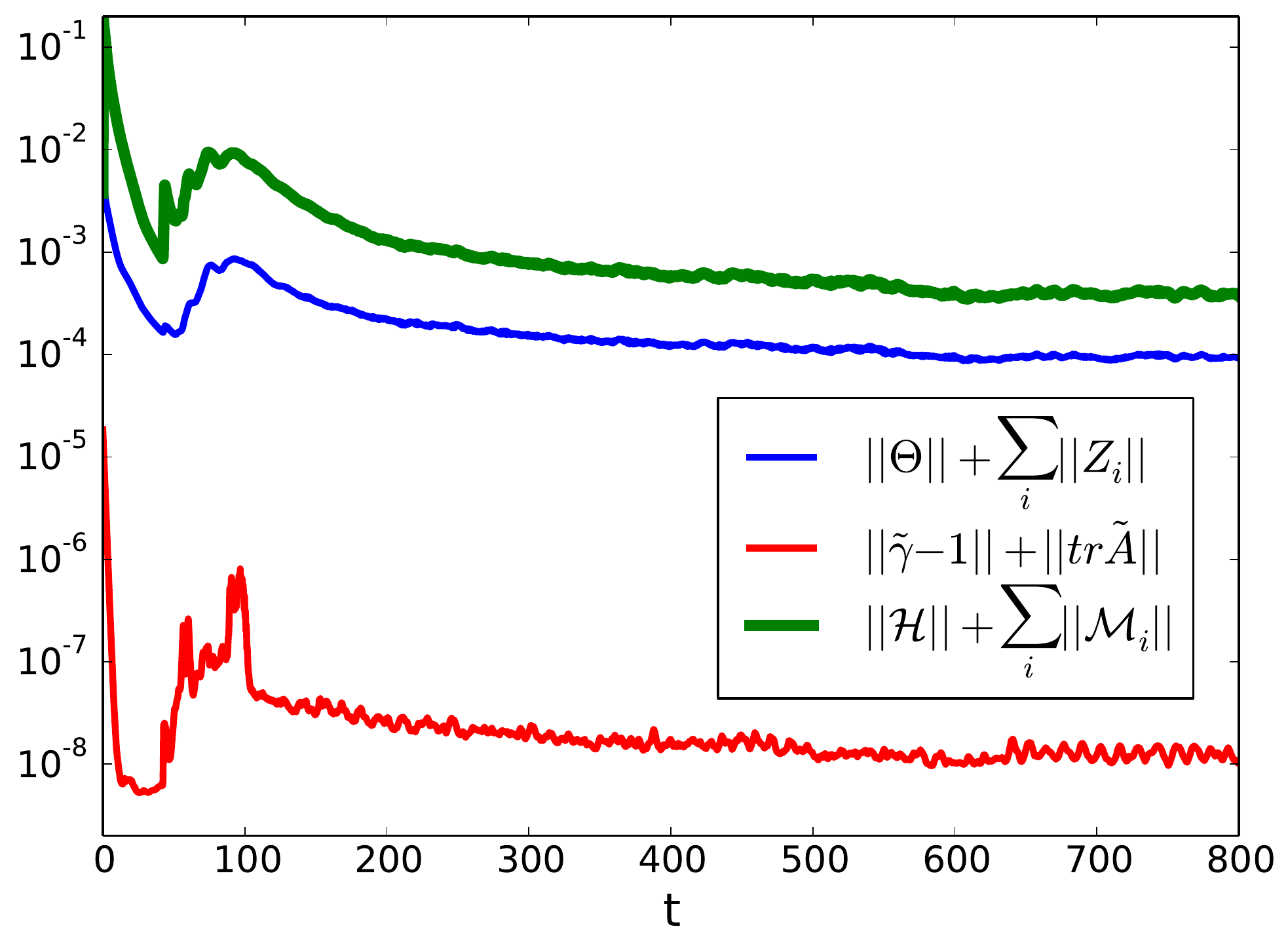}
\caption{{\em Orbital binary collisions.} $L_{2}$-norm of physical (blue solid line), the conformal (redd solid line) and the energy-momentum (green solid) constraints as a function of time. With our choice of the damping parameters, all the constraints are perfectly under control during all the simulation.}
\label{l2norm_orb}
\end{figure}

Finally, the evolution of some constraints is displayed in Fig.~\ref{l2norm_orb}, showing that they are small and kept under control during all the simulation.

% % % % % % % % % % % % % % % % % % % % % % % % % %
% % % % % % % % % % % % % % % % % % % % % % % % % %
\section{Conclusions}\label{con}
% % % % % % % % % % % % % % % % % % % % % % % % % %
% % % % % % % % % % % % % % % % % % % % % % % % % %

In this work we have studied the merger of binary compact boson stars, focusing on two different cases: head-on collisions of boson-boson/boson-antiboson for a wide range of shift phases, and orbital mergers of identical stars with different velocities. 

These simulations have been performed by solving the Einstein-Klein-Gordon system with a novel modification of the CCZ4 formalism which does not require the algebraic enforcement of the conformal constraints after each step of the numerical integration. Our formulation treats these conformal constraints in the same way than the physical ones, and they are also kept under control by including damping terms. We have studied the pseudo-hyperbolicity of the evolution system using a linear plane-wave analysis to show that, only for the preferred choice $\lambda_{0}=1$, the system has a complete set of eigenvectors with real eigenvalues. Therefore, the linearized system is well posed, provided that $\lambda_{0}=1$. We have performed some standard numerical tests to check the robustness of our implementation, including the evolution of an isolated solitonic boson star. Our numerical results --stable evolutions with small and bounded constraints-- confirm the analytical analysis.

Within this formulation we have studied binary solitonic boson stars systems with and without angular momentum. All the binary systems considered were constituted by two equal-mass solitonic boson stars. In the head-on cases we allowed for non-identical boson stars with different rotation and shift phase. Our simulations show that the merger of a boson-boson binary leads, in general, to another solitonic boson stars. However, when the phase shift approaches $\pi$, the scalar field interaction is repulsive and stronger than gravity, preventing the merger of the two stars. The merger of a boson star and an anti-boson star completely annihilates each other for any of the phase shift considered. 
This behavior, combined with the results described in~\cite{pale1,pale2}, allow us to hypothesize that the  generic behavior during the collision of a boson and an antiboson star is the annihilation of both, independently on the interaction potential and the phase shift,   producing large amounts of unbound scalar field that is radiated to infinity.

In the scenario of orbital binaries, our studies with identical boson stars revealed that the merger always lead to the formation of a rotating bar which sheds quickly all its angular momentum by emitting scalar field and gravitational waves, to finally relax into a non-rotating boson star. This inability to form a rotating boson star from a merger might be due to the angular momentum quantization of the rotating solutions. Of particular interest is the case with the highest angular momentum considered, leading roughly to a system in quasi-circular orbits. In this case, soon after the merger, two blobs of scalar field, carrying away small amount of Noether charge but large fraction of angular momentum, were expelled from the remnant almost at light speed.     

Future studies will further study orbital binary systems
by considering different masses and analyzing the gravitational waves produced during the coalescence.

% % % % % % % % % % % % % % % % % % % % % % % % % % % %
% % % % % % % % % % % % % % % % % % % % % % % % % % % % 
\subsection*{Acknowledgements} 

It is a pleasure to thank Vitor Cardoso, Luis Lehner, Steve Liebling and Paolo Pani for interesting discussions on the subject and helpful comments on the manuscript, as well as to Hiro Okawa for useful suggestions on the initial data.
We acknowledge support from the Spanish Ministry of Economy and Competitiveness grants FPA2013-41042-P and AYA2016-80289-P (AEI/FEDER, UE). CP acknowledges support from the Spanish Ministry of Education and Science through a Ramon y Cajal grant. MB would like to thank CONICYT Becas Chile (Concurso Becas de Doctorado en el Extranjero) for financial support.
\section{Appendix}
\appendix

% % % % % % % % % % % % % % % % % % % % % % % % % % % %
% % % % % % % % % % % % % % % % % % % % % % % % % % % %

\section{Analysis quantities}\label{ap1}

Hence, 
the three dimensional Christoffel symbols  tensor can be written in function of conformal metric as:     
\begin{align}
\Gamma^i{}_{jk} = &\quad {\tilde \Gamma}^i{}_{jk} - {1 \over 2} \, \Bigl[ \partial_j \ln \chi \, {\tilde \gamma}^i{}_k + \partial_k \ln \chi \, {\tilde \gamma}^i{}_j \nonumber \\
&- \partial_m \ln \chi \, {\tilde \gamma}_{jk} \, {\tilde \gamma}^{im} \Bigr] 
\end{align}  
Besides, the BSSN conformal \cite{BSSN} connections is defined by  
\begin{equation}
{\tilde \Gamma}^i = {\tilde \gamma}^{jk} \, {\tilde \Gamma}^i{}_{jk} = - \partial_j {\tilde \gamma}^{ij}. 
\end{equation}

There are several analysis quantities in the CCZ4. The first term of the momentum constraint can be computed directly from this relation
\begin{align}
 \nabla_n {K^{n}}_{i} &=
   {\gamma}^{mn} \nabla_m K_{ni} \nonumber \\
    &=  {\tilde \gamma}^{mn} \partial_m {\tilde A}_{ni}
 - {\tilde \Gamma}^l{}_{mi} {\tilde A}^m{}_{l}
 - {\tilde \Gamma}^l {\tilde A}_{li}\nonumber \\
 & - \frac{3}{2 \chi} {\tilde A}^n{}_{i} \partial_n \chi
 + \frac{1}{3} \partial_i trK,
\end{align}

Assuming that $\theta=Z_i=0$, we have that Hamiltonian and momentum constraint are given by
\begin{eqnarray}
  {\cal H} &=& R + \frac{2}{3} (tr K )^2 - {\tilde A}_{ij} {\tilde A}^{ij} - 16 \pi G \rho 
\nonumber \\
  {\cal M}_i &=& 
  {\tilde \gamma}^{jk}  \partial_j  {\tilde A}_{ki} 
  - {{\tilde \Gamma}^j}_{ki} {\tilde A}^k_{j}
  - {\tilde \Gamma}^j {\tilde A}_{ij} \\ \nonumber
  & &- \frac{3}{2 \chi} {\tilde A}^j_i \partial_j \chi
  - \frac{2}{3} \partial_i trK
  - 8 \pi G \frac{{\tilde S}_i}{\chi}
\nonumber
\end{eqnarray}

Other interesting quantities are ADM mass and angular momentum, they can be computed by performing a surface integral at spatial infinity
\begin{eqnarray}
 M_{ADM} &\equiv& \frac{1}{16 \pi} \lim_{r\to\infty}
   \int_{S} \gamma^{ij} \left( \partial_j \gamma_{ik} - \partial_k \gamma_{ij} \right) dS^k
\nonumber \\
&=& \frac{1}{8 \pi} \lim_{r\to\infty}
   \int_{S} \left( {\tilde \gamma}^{ik} \partial_k \chi 
    + \frac{\chi}{2} {\tilde \Gamma}^{i} \right) dS_i
\label{admmass}
\\
 J^i_{ADM} &\equiv& \frac{1}{8 \pi} \lim_{r\to\infty}
   \int_{S} \epsilon^{ijk} x_j K_{kl} dS^l
\label{admJz}   
\\
&=& \frac{1}{8 \pi} \lim_{r\to\infty}
      \int_{S} {\hat \phi}_j \chi \left( {\tilde A}^{ji} + \frac{{\tilde \gamma}^{ji}}{3} trK\right) dS_i   
\nonumber      
\end{eqnarray}
in terms of the conformal quantities, 
where ${\hat \phi}_j = (-y, x, 0)$ is the cartesian coordinate basis axial vector. The surface element is given by
\begin{align}
  dS_i = \frac{x_i}{r} \chi^{-3/2} r^2 sin\theta d\theta d\phi
\end{align}

Other interesting analysis integral quantities, which are exact for stationary and stationary spacetimes, are the Komar mass and angular momentum~\cite{gour}:
\begin{eqnarray}
 M_{K} &\equiv& 2 \int_{\Sigma_{t}} \left(T_{ab}-\frac{1}{2} trT\,g_{ab}\right)
 n^{a} \zeta_t^{b} \sqrt{\gamma}\,d^{3}x
\nonumber \\
 J^z_{K} &\equiv& - \int_{\Sigma_{t}} \left(T_{ab}-\frac{1}{2} trT\,g_{ab}\right)
 n^{a} \zeta_{\varphi}^{b}\sqrt{\gamma}\,d^{3}x
\nonumber 
\end{eqnarray}
where $\zeta_t^{a}=(1,0,0,0)$ and ${\zeta}_{\varphi}^{a}=(0,{\hat \phi}^i/||{\hat \phi}^i||)$ are the time-like and the axial Killing vectors.

%\section{Gauge waves}
%Therefore, in ADM variables the initial data is given by
%\begin{eqnarray}
%  \alpha &=& \sqrt{H} ~~,~~ \gamma_{xx} = H ~~,~~ \gamma_{yy}=\gamma_{zz}=1 
%\\
%  K_{xx} &=& - \frac{1}{2 \alpha} \partial_t H = 
%  - \frac{1}{2 \sqrt{H}} A k \cos \left[ k x \right]
%\end{eqnarray}
%
%In terms of the CCZ4 variables the initial data is given by
%\begin{align}
%  \alpha &= \sqrt{H} ,\,\, \chi = H^{-1/3} ,\\
%   {\tilde \gamma}_{xx} &= H^{2/3} ,  {\tilde \gamma}_{yy}= {\tilde \gamma}_{zz}= \chi = H^{-1/3}\\
%  trK &= \gamma^{xx} K_{xx} = K_{xx}/\gamma_{xx} = K_{xx}/H\\  
%  {\tilde A}_{xx} &= \chi \left[K_{xx} - H trK/3 \right] ,\nonumber \\
%   {\tilde A}_{yy} &= {\tilde A}_{zz}= \chi \left[- H trK/3 \right]\\
%{\tilde \Gamma}^x &=  {\tilde \gamma}^{xx} {\tilde \gamma}^{xx} \partial_x {\tilde \gamma}_{xx} = 
%\frac{2}{3} H^{-5/3} \partial_x H        
%\end{align}

%\newpage

\bibliographystyle{utphys}
\bibliography{biblio}

\end{document}